\documentclass{article}
\usepackage[hyphens]{url}
\usepackage[utf8]{inputenc}
\usepackage[sort&compress, numbers]{natbib}
\usepackage{graphicx}
\usepackage{authblk}
\usepackage[english]{babel}
\usepackage{setspace}
\usepackage{fancyhdr}
\usepackage{array}
\usepackage[margin=1in]{geometry}
\usepackage{float}
\usepackage{datetime}
\usepackage{amsmath}
\usepackage{amssymb}
\usepackage{bbm}
\usepackage{bm}
\usepackage[final]{pdfpages}
\usepackage{longtable}
\usepackage{chngcntr}
\usepackage{placeins}
\usepackage{microtype}
\usepackage{tikz}
\usepackage{makecell}
\usepackage{hyperref}
\usepackage{subcaption}
\usepackage{rotating}
\usepackage{soul}
\usepackage{tabularx}

\hypersetup{
    colorlinks = true,
    citecolor = {blue},
    urlcolor = {blue},
    menucolor = {blue},
    linkcolor = {blue}
}

\newcommand{\franz}[1]{\textcolor{black}{#1}}

\newcolumntype{L}[1]{>{\raggedright\let\newline\\\arraybackslash\hspace{0pt}}m{#1}}
\newcolumntype{C}[1]{>{\centering\let\newline\\\arraybackslash\hspace{0pt}}m{#1}}
\newcolumntype{R}[1]{>{\raggedleft\let\newline\\\arraybackslash\hspace{0pt}}m{#1}}

\makeatletter
\def\@author{}
\renewcommand\@author{\ifx\AB@affillist\AB@empty\AB@author\else
      \ifnum\value{affil}>\value{Maxaffil}\def\rlap##1{##1}%
    \AB@authlist \\*[0.1cm] \small on behalf of the EU-PEARL NASH Investigators\textsuperscript{+} \\[\affilsep]\AB@affillist
    \else  \AB@authors\fi\fi}
\makeatother

\title{Designing an exploratory phase 2b platform trial in NASH with correlated, co-primary binary endpoints}

\author[1]{Elias Laurin Meyer}
\author[2]{Peter Mesenbrink}
\author[3]{Nicholas A. Di Prospero}
\author[4,5]{Juan M. Pericàs}
\author[6,7]{Ekkehard Glimm}
\author[8]{Vlad Ratziu}
\author[4]{Elena Sena}
\author[1,*]{Franz König}
\affil[1]{Center for Medical Data Science, Medical University of Vienna, Austria}
\affil[2]{Novartis Pharmaceuticals Corporation, One Health Plaza, East Hanover, NJ, USA}
\affil[3]{Janssen Research and Development, Raritan, NJ, USA}
\affil[4]{Liver Unit, Internal Medicine Department, Vall d'Hebron University Hospital, Vall d'Hebron Institute for Research (VHIR), 08036 Barcelona, Spain}
\affil[5]{Centros de Investigación Biomédica en Red Enfermedades Hepáticas y Digestivas (CIBERehd), ISCIII, Madrid, Spain}
\affil[6]{Novartis Pharma AG, Basel, Switzerland}
\affil[7]{Institute of Biometry and Medical Informatics, University of Magdeburg, Germany}
\affil[8]{Assistance Publique-Hôpitaux de Paris, Hôpital Pitie-Salpetriere, University of Paris, Paris, France}
\affil[+]{The list of investigators is shown in the Acknowledgements}
\affil[*]{Correspondence: franz.koenig@meduniwien.ac.at; Tel.: +43-1-40400-74800}
\date{}         
\begin{document}

\maketitle

\section*{Abstract}

Non-alcoholic steatohepatitis (NASH) is the progressive form of nonalcoholic fatty liver disease (NAFLD) and a disease with high unmet medical need. Platform trials provide great benefits for sponsors and trial participants in terms of accelerating drug development programs. In this article, we describe some of the activities of the EU-PEARL consortium (EU Patient-cEntric clinicAl tRial pLatforms) regarding the use of platform trials in NASH, in particular the proposed trial design, decision rules and simulation results. For a set of assumptions, we present the results of a simulation study recently discussed with two health authorities and the learnings from these meetings from a trial design perspective. Since the proposed design uses co-primary binary endpoints, we furthermore discuss the different options and practical considerations for simulating correlated binary endpoints. 

\section{Introduction} \label{sec:intro}

The recent years have seen unprecedented challenges for many branches of modern medical research. The desire to accelerate development and approval of new treatments has called into question some long-standing drug development paradigms, such as the strict succession of phase 1, 2 and 3 trials and the insistence on separate trials for every experimental compound \citep{Redman2015}. Consequently, substantial effort has been made into the development of master protocol trials and in particular platform trials \citep{Meyer2020, Meyer2021, Woodcock2017, phdthesis}. These types of trials allow evaluation of many investigational treatments in parallel and hence their implementation has increased over the last years. The interest in platform trials has increased further with the emergence of the global pandemic due to the SARS-CoV-2 virus \citep{Kunz2020, Stallard2020Covid, Dodd2020, Horby2020, Angus2021, Macleod2021}. However, many operational, logistical and statistical challenges around platform trials remain. \newline
\newline
The definition of platform trials used in this article is that they are clinical trials which investigate multiple treatments or treatment combinations in the context of a single disease, possibly within several sub-studies for different disease sub-types or targeting different trial participant populations. In a platform trial, both drugs or drug combinations within existing sub-studies, as well as new sub-studies, may enter or leave the trial over time, allowing the trial to run infinitely, in principle. Within each sub-study, many adaptive and innovative design elements may be combined that clearly separate platform trials from more classical trial designs \citep{Woodcock2017}. For a more detailed introduction, we refer to \citet{Meyer2020}, where we conducted a comprehensive systematic search to review current literature on master protocol trials from a design and analysis perspective. \franz{A compact glossary of common terms related to platform trials can be found in Table \ref{tab:glossary}, while a more detailed list of terms and explanations can be found online \citep{D2.1}.} \newline
\newline
Platform trials can leverage their main strengths such as adaptive design elements, testing multiple hypotheses in a single trial framework, reduced time to make decisions, ease of incorporating new investigational treatments into the ongoing trial and possibilities for collaboration between different consortia/sponsors. In 2018, the Innovative Medicines Initiative (IMI) put forth a call for proposals for the development of integrated research platforms to conduct platform trials to enable more patient-centric drug development. A consortium of 36 private and public partners have come together in a strategic partnership to deliver on the IMI proposal goals; the project is called EU Patient-cEntric clinicAl tRial pLatforms (EU-PEARL) \citep{Pearl}. Among the expected outputs of the initiative are publicly available master protocol templates for platform trials and four disease-specific master protocols for platform trials ready to operate in disease areas still facing high unmet clinical need; one of those diseases being non-alcoholic steatohepatitis (NASH). \newline
\newline
NASH is a more progressive form of non-alcoholic fatty liver disease (NAFLD) and is estimated to affect approximately 5\% of the world population. The disease is characterized by the accumulation of fat in the liver in the absence of significant alcohol intake or other secondary causes of hepatic steatosis \citep{chalasani2018diagnosis, powell2021non}. Over time, chronic inflammation and liver cell injury  lead to fibrosis and eventually cirrhosis including complications of end-stage liver disease and hepatocellular carcinoma. Indeed, NASH complications are rapidly becoming the leading indication for liver transplantation. In addition, NASH is associated with higher risks of developing cardiovascular diseases, which is the primary cause of death for most people affected. Currently, there are no approved treatments for NASH in the US and EU and in recent years several compounds failed to meet their phase 3 primary endpoint(s) \citep{ratziu2020so, ratziu2022breakthroughs}. However, developing treatments for NASH is a very active area of clinical research with dozens of industry-sponsored interventional studies active or recruiting trial participants \franz{across phases 1 through 3 with the vast majority in phase 1 or 2} according to ClinialTrials.gov and the EU clinical trials register (https://www.clinicaltrialsregister.eu). \newline   
\newline
\franz{To facilitate and accelerate the identification of the most effective and promising novel treatment options for trial participants with NASH, multiple potential novel therapies, as well as combinations of novel mechanisms of action, will need to be evaluated in well-designed early clinical studies before advancing to pivotal phase 3 programs. From a platform study perspective, Phase 2b is often the preferred trial design as it generally offers a robust pipeline for most indications and the ability to make decisions more rapidly before committing to longer, more costly development. This is particularly true for NASH where there is an abundance of compounds in early development and phase 3 programs tend to run over several years. Importantly, there are broadly common design elements, study populations, procedures, and endpoints for NASH phase 2b clinical studies which are aligned with Health Authority (HA) guidance. \newline  
\newline
Both the United States Food and Drug Administration (FDA) and the European Medicines Agency (EMA) have put forward advice for developing drugs for patients with non-cirrhotic NASH \citep{fda_nash, ema_nash, anania2020nonalcoholic}. Both HAs  note that the risk of progression to clinical outcomes (i.e., both liver-related and non liver-related morbidity and mortality) is mainly related to fibrosis stage.  Therefore, the non-cirrhotic NASH population that should be studied are individuals with either fibrosis stage 2 (F2) or stage 3 (F3) since they are at increased risk of progression relative to those with little (F1) or no (F0) liver fibrosis \citep{hagstrom2017fibrosis, dulai2017increased, ratziu2018critical}. In addition, the recognition by the HAs that the length of time necessary to observe a sufficient number of clinical events to assess drug efficacy may hamper drug development has led the HAs to recommend improvement in liver histology as clinical trial endpoints (i.e., resolution of steatohepatitis and no worsening of liver fibrosis, improvement in liver fibrosis greater than or equal to one stage with no worsening of steatohepatitis), which can be used as surrogates for approval in Phase 3 according to the accelerated approval pathways. Therefore, the FDA guidance advises that phase 2b studies demonstrate efficacy on a histological endpoint after at least 12-18 months of treatment, given that histological change takes an extended period of time to occur using a range of doses to support phase 3 dose selection. Therefore, members of EU-PEARL are currently developing a master protocol to support a phase 2b platform trial in NASH and this paper, as well as a previously published simulation study \citep{Meyer2022_1}, describe the initiative’s efforts to simulate the performance of the parameters used to make decisions on whether or not the treatment being evaluated is effective.}

\begin{longtable}{@{\extracolsep{5pt}} L{4cm}L{10cm}} 
\caption{Glossary for important terms related to platform trials, taken partly from ICH E9 \citep{ich1999statistical}, partly EU-PEARL D2.1 \citep{D2.1}.}

  \label{tab:glossary}
  \\[-1.8ex]\hline 
\hline \\[-1.8ex] 
Term & Description \\
\\[-1.8ex]\hline 
\hline \\[-1.8ex] 
Adaptive Design & An adaptive design allows the pre-specification of flexible components to the major aspects of the trial, like the treatment arms used (dose, frequency, duration, combinations, etc.), the allocation to the different treatment arms, the eligible patient population, and the sample size. An adaptive design can learn from the accruing data what the most therapeutic doses or arms are, allowing for example, the design to home in on the best arms. \\
\\[-1.8ex]\hline 
Integrated Research Platform & An Integrated Research Platform (IRP) is a novel clinical development concept centered on a master trial protocol which can accommodate multi-sourced interventions using the existing infrastructure of hospitals and federated patient data in design, planning and execution, while an optimized regulatory pathway for these novel treatments has been assured. \\
\\[-1.8ex]\hline 
Master Protocol & The term “master protocol” refers to a single overarching design developed to evaluate multiple hypothesis, and the general goals are to improve efficiency and establish uniformity through standardization of procedures in the development and evaluation of different interventions. Under a common infrastructure, the master protocol may be differentiated into multiple parallel sub-studies to include standardized trial operational structures, patient recruitment and selection, data collection, analysis, and management. In a platform trial the protocol will have the infrastructure to drop interventions and allow new interventions or combinations of interventions to enter
the study based on decision rules in the master protocol. \\
\\[-1.8ex]\hline 
Platform Trials & Clinical trials which investigate multiple treatments or treatment combinations in the context of a single disease, possibly within several sub-studies for different disease sub-types or targeting different trial participant populations. For more information, see section \ref{sec:intro}. \\
\\[-1.8ex]\hline 
Multi-center Trial & A clinical trial conducted according to a single protocol but at more than one site, and therefore, carried out by more than one investigator. \\
\\[-1.8ex]\hline 
Frequentist Methods & Statistical methods, such as significance tests and confidence intervals, which can be interpreted in terms of the frequency of certain outcomes occurring in hypothetical repeated realisations of the same experimental situation. \\
\\[-1.8ex]\hline 
Bayesian Methods & Approaches to data analysis that provide a posterior probability distribution for some parameter (e.g. treatment effect), derived from the observed data and a prior probability distribution for the parameter. The posterior distribution is then used as the basis for statistical inference. \\
\\[-1.8ex]\hline 
Interim Analysis & Any analysis intended to compare treatment arms with respect to efficacy or safety at any time prior to the formal completion of a trial. \\

\hline

\end{longtable}

\section{Methods}

\subsection{Platform Design}

\begin{figure}[ht]
\centering
\includegraphics[scale=0.37]{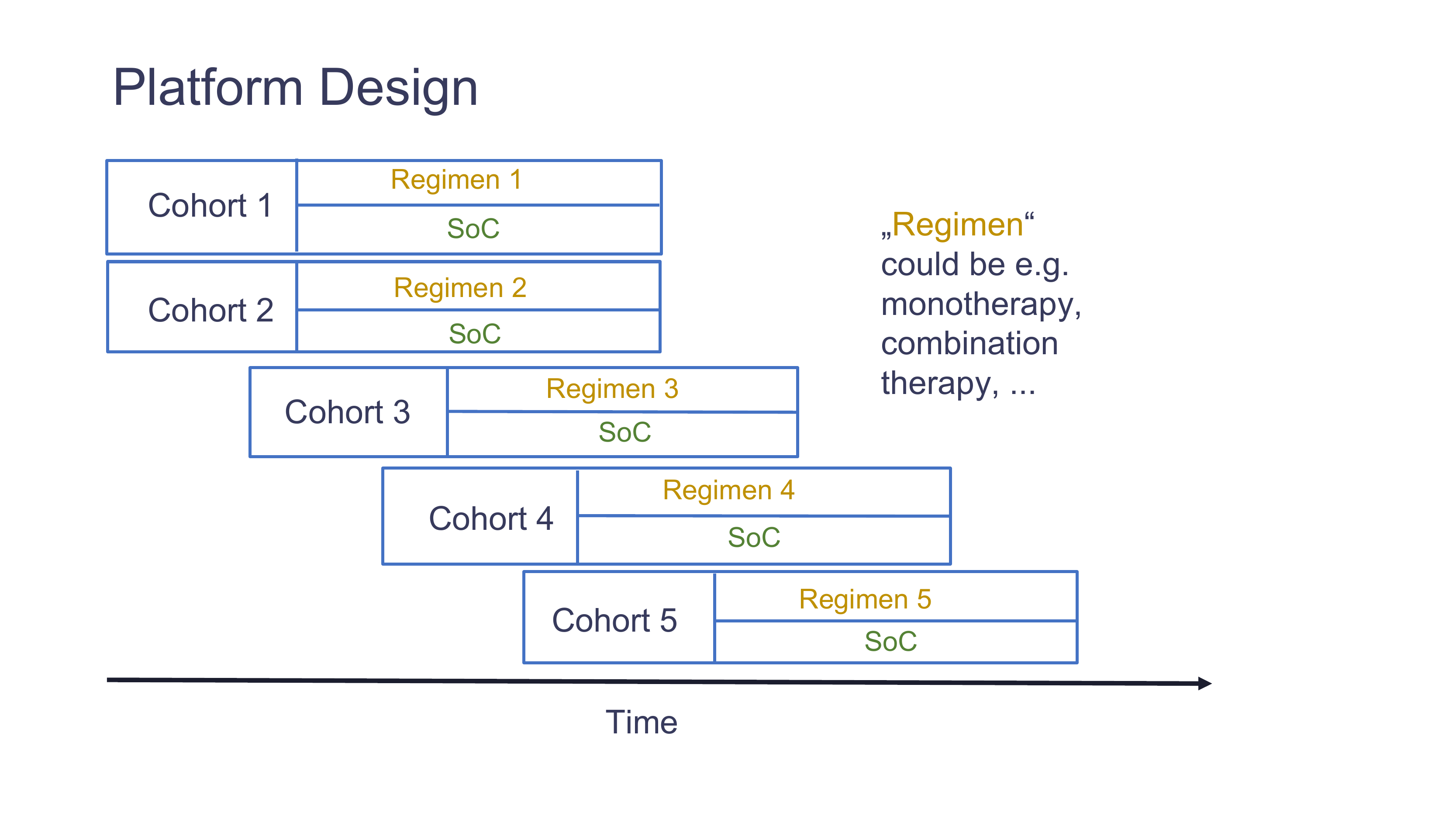}
\caption{Phase 2b platform trial design in non-alcoholic steatohepatitis (NASH). After an initial inclusion of two cohorts consisting of control (usually the standard-of-care, "SOC") and "regimen" arm (which could be a monotherapy or a combination therapy), more cohorts of the same structure are entering the trial over time. Within each cohort, several interim and a final analysis are conducted using the co-primary binary endpoints "NASH resolution without worsening of fibrosis" and "Fibrosis improvement without worsening of NASH". The platform trial ends when all cohorts have been evaluated.}
\label{fig:wp6}
\end{figure}

An overview of the proposed platform trial design can be found in Figure \ref{fig:wp6}. Generally, it is assumed that after an initial inclusion of a certain number of cohorts each consisting of treatment and matching control, further cohorts will enter over time while some of the existing cohorts might be discontinued for efficacy or futility. Trial participants entering the platform will be allocated between open cohorts. Within open cohorts, trial participants will be equally allocated between control and treatment arm using a block randomization of length two. Finally, the platform ends when all cohorts have finished their analyses. If the inclusion and exclusion criteria of the different cohorts are similar, it might be preferable to share the accumulating information on the control treatments, at least for concurrently enrolling trial participants. While there is a lot of controversy regarding the use of non-concurrent controls \citep{Bofill2021}, sharing only information on trial participants that could have been randomized to the arm under investigation seems uncontroversial (note that this requires data to be concurrent). As noted before, platform trials can run perpetually without limiting the number of drugs going into the trial. Any potentially successful compound in a NASH phase 2b trial would have to show either resolution of NASH without worsening of fibrosis (binary endpoint 1) and/or 1-stage fibrosis improvement without worsening of NASH (binary endpoint 2).  \newline
\newline
Endpoints 1 and 2 are correlated binary endpoints and clinical studies have demonstrated a strong link between histologic resolution of steatohepatitis with improvement in fibrosis \citep{brunt2019improvements, kleiner2019association}, therefore, improvement in endpoint 1 could lead to improvement in endpoint 2 but not necessarily the converse \franz{and not necessarily during the same time frame. The current regulatory guidance is that the FDA recommends demonstrating endpoint 1 OR endpoint 2 and the EMA recommends demonstrating endpoint 1 AND endpoint 2 \citep{hagstrom2017fibrosis, dulai2017increased, ratziu2018critical}}. For this simulation study, we decided to follow FDA endpoint recommendations. Within EU-PEARL, several possible phase 2b platform trial designs for NASH were considered - treatment (one dose; could be monotherapy or combination therapy) versus control, treatment (multiple doses; could be monotherapy or combination therapy) versus control, combination therapy versus monotherapies versus control, etc. Furthermore, it was considered whether the final endpoints (which are observed after roughly 48-52 weeks) should be used for interim decision making or whether a short-term surrogate endpoint should be used. Based on the proposed design, comprehensive simulations were run for two scenarios: monotherapy (one dose) versus control and combination therapy versus monotherapies versus control. We will present results of the former in this paper and results of the latter can be found in \citet{Meyer2022_1, Meyer2022_2}.

\subsection{Decision Rules} \label{DR}

\begin{sidewaysfigure}[ht]
\centering
\includegraphics[scale=0.63]{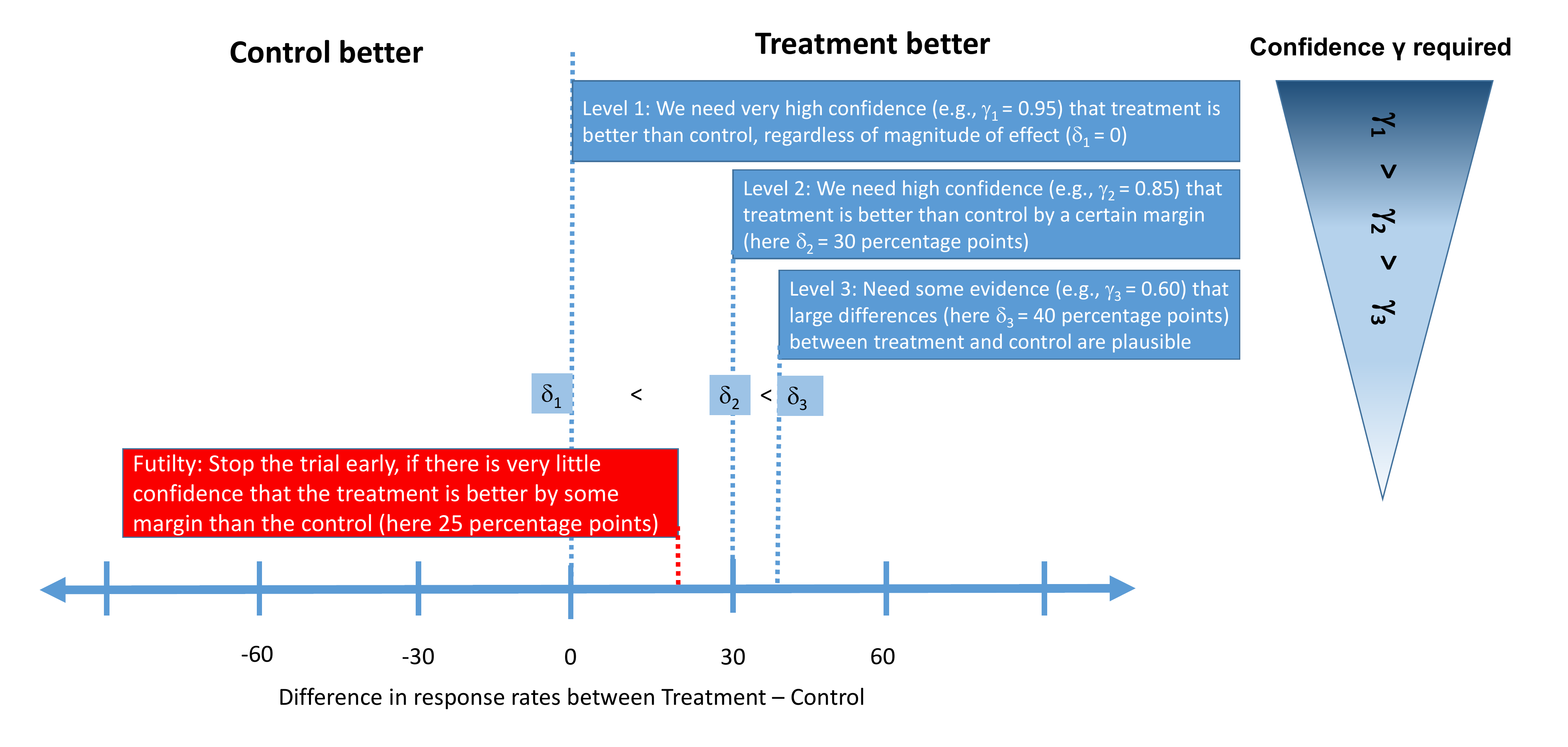}
\caption{\franz{Schematic overview of decision rules used. On the x-axis, the difference in response rates between the control and treatment group (i.e. treatment effect) in percentage points is shown. At the two interim analyses, cohorts can be stopped early for futility, if there is very little evidence (interim analysis 1: less than 20\%, interim analysis 2: less than 30\%) that the treatment is better than control by 25 percentage points or more (red box). At all analysis time points, the same efficacy decision rules are used (blue boxes). Depending on the aim of the study, all or only certain levels of evidence could be required (see also Table \ref{tab:hierarchy}). The treatment effects ($\delta$s) presented in this figure correspond to the decision rules used for endpoint 1 - for endpoint 2, we used $\delta_1 = 0, \delta_2 = 0.175, \delta_3 = 0.25$, as well as a futility margin of 10 percentage points.}}
\label{fig:dr}
\end{sidewaysfigure}

\franz{Decisions on whether or not to promote treatments to the next stage of development can be based on different principles such as fixed thresholds for treatment effect estimates, the p-values of statistical frequentist tests for treatment efficacy, conditional or predictive probabilities of final trial success. Many readers might be familiar with group-sequential trials where early stopping for futility or efficacy is based on the p-values from statistical tests which are adjusted for repeated looks into the data, such as the O’Brien-Fleming test \citep{jennison1999, sonja}. In some simple situations (e.g. if stopping the entire clinical trial for efficacy or futility is the only permitted interim decision option), it is possible to convert such decision rules into each other \citep{Gallo2014} (in the sense that a decision rule given by a threshold on conditional power can equivalently be stated by a correspondingly recalculated threshold on the estimated treatment effect, say). In platform trials, however, the decision space is usually more complicated and comprises interdependent decisions such as stopping arms without stopping the entire trial or selecting treatments if they are sufficiently superior to other treatments. In such situations, there is no simple 1-to-1 correspondence between decision rules formulated on different scales (e.g. a decision rule which is influenced by the treatment effect estimates from several treatments cannot be converted into a fixed threshold for one specific treatment). It is also very difficult to provide decision rules on un-standardized measures such as treatment effect estimates, since these would have to be derived anew for every concrete application. For these reasons, we focus on Bayesian posterior probabilities \citep{phdthesis} as the main vehicle for making decisions in this paper. The benefit of using Bayesian decision rules is their flexibility regarding extensions to several criteria and interim analyses}. To illustrate the basic mechanics, we introduce the concept for comparing the response rate of a new treatment ($\pi_E$) with the response rate of the Standard-of-Care (SoC) ($\pi_S$) in a clinical trial. For an analysis after observing data $D$, we are conducting a Bayesian analysis with the aim of deciding whether there is enough evidence to declare the treatment effective. First, we will introduce the concept for a Bayesian decision rule testing a single endpoint using a parsimonious notation for illustrative purposes. Later and in the appendix, we will show how the parameters could be specified for the endpoints at hand in NASH. Different levels of evidence will be introduced depending on the parameterization of the Bayesian decision rule. Typically, a Bayesian decision rule of the following sort could be used for comparing the new treatment to the control SoC (the priors on $\pi_S$ and $\pi_E$ are omitted for better readability):

\begin{align}
\begin{split}
\label{eq:DR1}
\text{Declare Efficacy, if } & P(\pi_{E} > \pi_{S} + \delta | D) > \gamma \\
\end{split}
\end{align} 
\newline
with some pre-specified probability threshold $\gamma$ and pre-defined margin $\delta$ for the targeted treatment effect of interest. Such a decision rule based on a posterior distribution can, but does not have to, correspond to a particular null hypothesis (e.g. $H_0: \pi_E > \pi_S + \delta$). \franz{For example, it is sometimes appropriate to use so-called "shrinkage estimators" where the single treatment effect estimates in a platform trial are "shrunk" towards a common average effect. This is appropriate if drugs share a common mechanism of action and it is therefore a priori plausible that they may have similar effects. In such situations, the decision on a single drug is influenced by the performance of the entire class of drugs.}  For better readability, the dependence on the data $D$ is omitted in following sections. \newline
\newline
Decision rules should provide a high level of confidence that graduating compounds are competitive with respect to the current landscape of compounds in development with publicly accessible phase 2/3 studies published. In particular, semaglutide demonstrated a 42 percentage point response rate increase in NASH resolution (endpoint 1) as compared to placebo \citep{newsome2021placebo} and lanifibranor demonstrated a 19 percentage point response rate increase in fibrosis improvement (endpoint 2) as compared to placebo \citep{francque2021randomized}.  \franz{In the process of eliciting which exact decision rules to use, we first conducted a review of studies in NASH. Several structured discussions were held between statisticians and clinical experts to define endpoints and targeted effect sizes. Finally, the experts provided confidence intervals based on which they would accept graduation of compounds from the platform trial. Confidence intervals are generally centered around the observed response rate and the width of the confidence interval describes the remaining uncertainty, which is directly linked to the sample size, i.e. larger sample sizes lead to narrower confidence intervals. The width of the confidence intervals the experts were presented with corresponds to a sample size of 75, which is the lowest treatment group size investigated in this simulation study and corresponds to a treatment group size usually used in NASH phase 2b trials.} In frequentist decision making, one might tailor the decision rules such that the confidence interval does not include a certain lower bound of efficacy. The multi-component Bayesian decision rules we propose will allow for refined specification of evidence on the efficacy of a new compound required in order to graduate, while at the same time controlling basic type 1 error with one of its decision rule components. It should also be noted that our efficacy decision rules are based on checking the criterion that there is sufficient confidence that the effect size (i.e. the difference in response rate between the experimental and control treatment) exceeds (a) certain margin(s) with certain confidence(s), i.e., the posterior probabilities for decision rules as defined in equation \ref{eq:DR1}. If the margin is selected close to the true (but unknown) effect size, there are limits for the achievable confidence, which in some situations seems counter-intuitive. As an example, if the true success rate is 0.5 and assuming a weakly informative prior, we will never be able to achieve a confidence greater than 50\% that the true success rate is 0.5 or larger (for large sample sizes). \franz{The reason is that the posterior distribution will be centered around the true success rate of 0.5, i.e., resulting in a probability of maximum 50 \% that the value will be equal or larger then 0.5.  This is equivalent to achieving 50\% power to detect a success rate of 0.5 if we require a confidence greater or equal to 50\%}. If indeed we wanted to detect a success rate of 50\% with a larger power, we need to either reduce the required confidence or the targeted success rate in our Bayesian decision rules. This is illustrated further in Figure \ref{fig:beta} in appendix \ref{sec:appBeta}, where the resulting posterior distribution for a theoretical success rate of 0.5 is shown if a weakly informative Beta (1,1) prior and a sample size of 75 participants per group are chosen and the observed success rate equals the assumed response rate. Finally, communication of the chosen decision rules to clinicians and general audiences is not always straightforward, e.g. while graduating a compound if there is 20\% confidence that the effect is sufficiently large (say $\Delta$) is equivalent to dropping the compound if there is more than 80\% confidence that the effect is smaller than $\Delta$, the latter is much more generally understood. For the lack of a comparable frequentist design, no direct comparison of operating characteristics was conducted. \newline
\newline
We propose a Bayesian framework for a multi-level efficacy decision rule which incorporates different levels of evidence, ranging from information whether the treatment is simply superior to the control up to information on how likely larger effect sizes of interest are. A specification for such multi-level efficacy decision rules using three levels of evidence can be found for both endpoints of interest in NASH in Table \ref{tab:hierarchy} and Figure \ref{fig:dr}. At level 1, the main target is to show whether the experimental treatment is superior to the control by setting the margin $\delta_1 = 0$ . To ensure sufficient type 1 error control (assuming weakly informative priors on the success rates) the required confidence is set to $\gamma_1 = 0.95$. At level 2, there should be sufficient evidence provided that the true effect size is larger than moderate differences with a certain level of confidence. For example, we set $\delta_2$ for each endpoint to an effect size which we elicited should be close to the lower end of any 95\% confidence interval based on which a treatment would be graduated and the required confidence, $\gamma_2$, to see an effect size at least as large as this to 85\% (in accordance with our considerations regarding the confidence interval). Level 3 requires that there is also sufficient confidence in observing larger effects. For example we set $\delta_3$ for each endpoint to an effect size which we elicited should be slightly below the center of any 95\% confidence interval based on which a treatment would be graduated and the required confidence, $\gamma_3$, to see an effect size at least as large is this to 60\% (again in accordance with our considerations regarding the confidence interval). Of course any such ordering of $\delta$s and $\gamma$s should fulfill the conditions $\delta_1 < \delta_2 < \delta_3$ and $\gamma_1 > \gamma_2 > \gamma_3$ to be meaningful. To motivate the multi-level decision rules, we simulated all scenarios using the different levels of required evidence. Please note that while $\delta_1 < \delta_2 < \delta_3$ and $\gamma_1 > \gamma_2 > \gamma_3$, this does not mean that level 3 decision rules are generally "stricter" than level 2 or level 1 decision rules. In fact, if the posterior was extremely flat, the level 1 requirement would be the strongest.

\begin{longtable}{@{\extracolsep{5pt}} L{3cm}L{2cm}L{2cm}L{7cm}} 
\caption{Different levels of evidence required to graduate treatment for efficacy. The ordering is hierarchical in nature, i.e. requiring two levels of evidence means level 1 and level 2 need to be simultaneously fulfilled. E1 and E2 refer to endpoint 1 (resolution of NASH without worsening of fibrosis) and endpoint 2 (1-stage fibrosis improvement without worsening of NASH) respectively.}

  \label{tab:hierarchy}
  \\[-1.8ex]\hline 
\hline \\[-1.8ex] 
Level of Evidence $l$ & Margin $\delta$ for targeted difference & Confidence $\gamma$ required & Description \\
\\[-1.8ex]\hline 
\hline \\[-1.8ex] 

1  & 0 (both endpoints) & 95\% & First level of efficacy evidence required serving as a threshold to establish sufficient confidence in any treatment effect larger than 0, i.e. superiority of a treatment to control. Note that when using non-informative priors, the one-sided frequentist type I error will be about $1-\gamma$ for a single endpoint, i.e. 5\% in this example.  \\
\hline

2  & 0.30 (E1) 0.175 (E2) & 85\% & Second level of efficacy evidence required serving as a threshold to establish a high confidence that the true effect is larger than moderate treatment effects. In the example a higher margin for the first endpoint is required compared to the second endpoint. \\
\hline

3 & 0.40 (E1) 0.25 (E2) & 60\% & Third level of efficacy evidence required serving as a threshold to establish sufficient confidence in large treatment effects. \\
\hline

\end{longtable}

In addition to graduating a treatment based on (possibly multi-level) Bayesian decision rules, one might also be interested in dropping a treatment at an interim analysis for futility based on posterior probabilities. In this case, the Bayesian decision rule for a single endpoint can be expanded introducing margins $\delta_E^{T}$ and $\delta_{F}^{T}$ with thresholds $\gamma_E^{T}$ and $\gamma_F^{T}$ for graduating and dropping, respectively. At a given point in time $T$, after observing data $D$, we are conducting an analysis with the aim of deciding whether we have enough evidence to declare the treatment efficacious or futile.
\begin{align}
\begin{split}
\label{eq:DR}
\text{Declare Efficacy, if } & P(\pi_{E} > \pi_{S} + \delta_E^{T} | D) > \gamma_E^{T} \\
\text{Declare Futility, if } & P(\pi_{E} > \pi_{S} + \delta_{F}^{T} | D) <  \gamma_F^{T} \\
\text{Continue Trial, } & \text{otherwise}
\end{split}
\end{align} 

with some pre-specified probability thresholds $\gamma_E^{T}$ and $\gamma_{F}^{T}$ and required treatment effects $\delta_E^{T}$ and $\delta_F^{T}$. \franz{Since we decided to follow FDA requirements, a treatment will be graduated if the efficacy decision rules are met for at least one of the two endpoints (this will be referred to as the "OR" decision rule), while it will only be dropped at interim for futility, if the futility decision rules are met for both of the endpoints. The specification of the efficacy decision rules for both endpoints is given in Table \ref{tab:hierarchy}, whereby the same margins $\delta$ and confidence $\gamma$ are used for the interim and final analyses. The trial will be stopped for futility, if there is only a small likelihood that the response rates of the treatment arm exceeds the control arm by at least 25 and 10 percentage points for endpoint 1 and 2, respectively. For endpoint 1, both efficacy and futility decision rules are illustrated in Figure \ref{fig:dr}.  For more information, including a verbal description of the decision rules and a more formal definition, please refer to appendix \ref{sec:app1}}.

\section{Simulations}

\subsection{Simulation Setup}\label{sim_setup}

\begin{figure}[ht]
\centering
\includegraphics[scale=0.44]{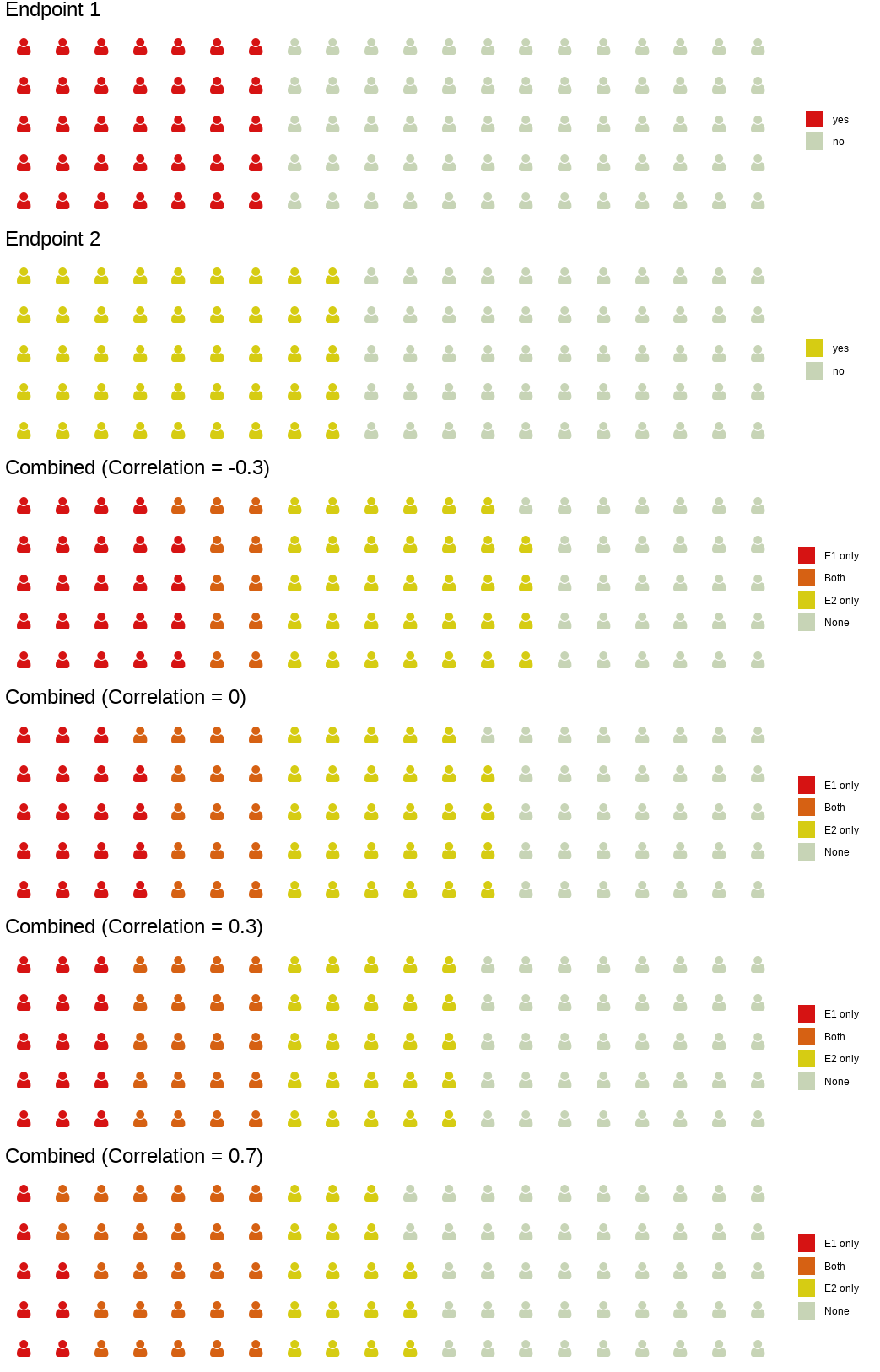}
\caption{\franz{Impact of different levels of correlation between endpoints 1 and 2 on the expected number of responders. Assuming a response rate of 30\% for endpoint 1 and 40\% for endpoint 2, we expect 30/100 patients to reach endpoint 1 (in red) and 40/100 patients to reach endpoint 2 (in yellow), regardless of the correlation. Depending on different levels of the correlation, the number of responders that reach both endpoints simultaneously (in orange) varies; it increases with increasing correlation. In contrast, the expected number of trial participants that reach at least on of the two endpoints decreases with increasing correlation (in this example, this number is 69 when the correlation is -0.3 and 53 when the correlation is 0.7).}}
\label{fig:corrsim}
\end{figure}

For classical randomized-controlled trials (RCTs) or even some multi-arm, multi-stage trials, the required sample size to achieve a certain power might be a deterministic function of several assumptions and design parameters such as treatment effects, significance level and chosen test procedure. Due to the complex designs, platform trials usually require simulations to be run in order to calculate operating characteristics such as power and average trial duration. In order to simulate the EU-PEARL NASH phase 2b platform trial, we chose a set of assumptions and design choices that were fixed and a set of assumptions and design choices that were varied (see Table \ref{tab:simsetup}). In general, for every trial participant we observe two correlated binary outcomes, NASH resolution (Endpoint 1) and fibrosis improvement (Endpoint 2). Outcomes are simulated using the approach described in section \ref{sec:MNT}, such that in the simulations we can fix the success rates of endpoint 1 and endpoint 2 and their latent variable correlation $\rho$ (see Figure \ref{fig:corrsim}). \franz{The correlation between the two endpoints can be interpreted in such a way that if there is a positive correlation between the two endpoints, then there is an increased likelihood that either events are observed in both endpoints jointly or not at all. If there is a negative correlation, it means if an event is observed in one endpoint, then there is a larger likelihood that no event is observed for the other one. If the two endpoints are uncorrelated ($\rho=0$), knowing if an event was observed for one endpoint gives no information as to whether or not an event is observed for the second one.} Sample sizes reflect number of trial participants with complete observations (i.e. paired biopsies) by the time of final analysis. After this number of trial participants were enrolled in a given treatment arm, enrollment to this treatment arm stops. 

\begin{longtable}{@{\extracolsep{5pt}} L{2.5cm}L{1cm}L{2cm}L{8.5cm}} 
\caption{Specification of important simulation parameters. Values are either fixed or varied in different simulation scenarios. For different simulation parameters, we differentiate between parameters that are considered a design choice ("D") and parameters that are considered an assumption ("A") regarding the future course of the platform trial or treatment effects (see second column "Type").}

  \label{tab:simsetup}
  \\[-1.8ex]\hline 
\hline \\[-1.8ex] 
Name & Type & Investigated Values & Description \\
\\[-1.8ex]\hline 
\hline \\[-1.8ex] 

Timing of new cohorts & A & 24 & Number of weeks after which a new treatment enters the platform trial. \\
\hline

Accrual rate  & A & 6 & Number of participants entering the trial per week (approximation based on the number of participating centers and trial participants per center). \\
\hline

SoC responder rates  & A & 10\% (E1) 20\% (E2) & Success rates for endpoint 1 (E1) and endpoint 2 (E2) in the standard-of-care (SoC) arm. \\
\hline

Endpoint 1 Responder Rate  & A & Range from 0.10 to 0.55 & Assumed responder rate of the investigational treatments for endpoint 1 (NASH resolution without worsening of fibrosis). \\
\hline

Endpoint 2 Responder Rate  & A & Range from 0.20 to 0.55 & Assumed responder rate of the investigational treatments for endpoint 2 (fibrosis improvement without worsening of NASH). \\
\hline

Time trend  & A & 0 & Assumed drift in the outcome responder rates over time. Simulations were conducted assuming no such drift. \\
\hline

Correlation between endpoint 1 and endpoint 2 & A & -0.3, 0, 0.3, 0.7 & Assumed correlation between the latent continuous analogues to endpoint 1 and endpoint 2 (see section \ref{sec:MNT} for more details). While we assume the true correlation to be positive, a negative value was added in the simulation study in order to investigate a larger range of values. \\
\hline

Initial cohorts  & D & 2 & Number of cohorts the platform trial is initiated with. \\
\hline

Cohort limit  & D & 5 & Maximum number of cohorts that can enter the platform trial over time. \\
\hline

Outcome observation time  & D & 52 & Number of weeks after enrollment at which the primary outcome is observed. \\
\hline

Interim Analyses Timing  & D & 50\% (IA1) 75\% (IA2) & Timings of interim analyses relative to final planned sample size (counting observed outcomes). \\
\hline

Final Cohort Sample Size & D & 150, 250 & Number of trial participants after which final analysis in a cohort is conducted. These numbers correspond to sample sizes usually used in NASH phase 2b trials (i.e. 75/125 per arm). \\
\hline

Data Sharing & D & concurrent, cohort & Different methods of data sharing used at analyses, either using concurrent data ("concurrent") or not sharing at all ("cohort").  \\
\hline

Decision Rule & D & Specific rules & See section \ref{DR}, as well as Table \ref{tab:hierarchy} and Figure \ref{fig:dr} for more details on the decision rules used. For a formal definition, see appendix \ref{sec:appDR}.  \\
\hline

Evidence Level & D & 1,2,3 & Different levels of evidence required in the Bayesian decision rules (see section \ref{DR}). By default, the highest level of evidence is required (i.e. level 3).  \\
\hline

\end{longtable}

\noindent Simulations of this trial design were performed using the \textbf{cats} package, which is downloadable on Github (\url{https://github.com/el-meyer/cats}) and CRAN (\url{https://cloud.r-project.org/web/packages/cats/index.html}). For each of the distinct combinations of simulation parameters the platform trial was simulated 10000 times. Results of those 10000 simulated platform trial trajectories were summarized for each of the sets of simulation parameters and visualized using lattice plots \citep{meyer2022interactive}. In particular, we present the success probability (i.e. the probability for a particular drug to be declared superior to control; this is equivalent to type 1 error when the drug is in truth futile and power when the drug is in truth efficacious).

\clearpage

\subsection{Main Simulation Results}

\begin{sidewaysfigure}[ht]
\centering
\includegraphics[scale=0.47]{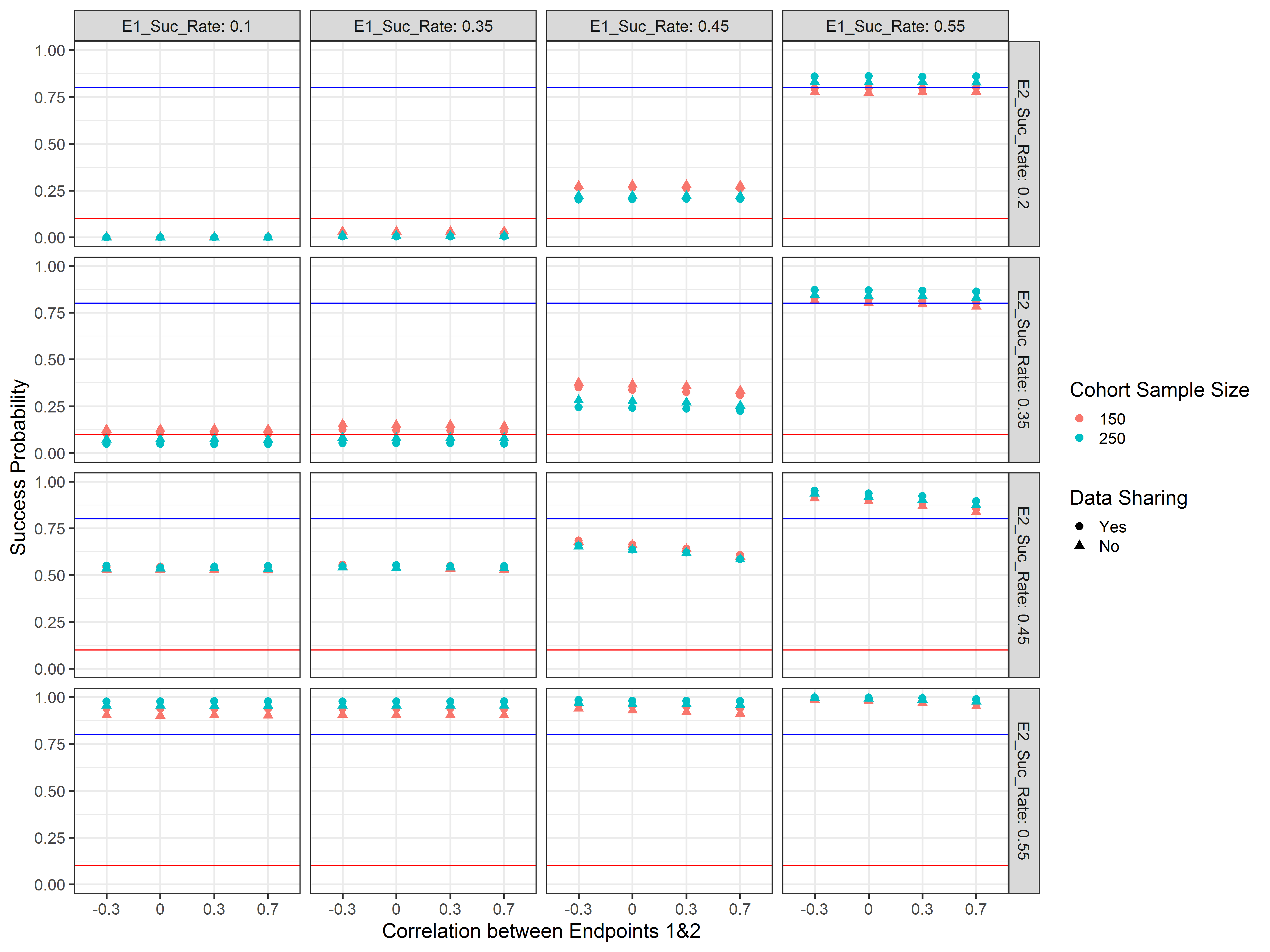}
\caption{Success probabilities for the treatment arm with respect to the response rate for endpoint 1 (E1\_Suc\_Rate; columns), the response rate for endpoint 2 (E2\_Suc\_Rate; rows), the correlation between the two endpoints (x-axis), the type of data sharing used (point shape) and the planned cohort sample size per arm (colour). The blue horizontal line marks 80\% as a common target for the power and the red horizontal line marks 10\% as a common target for type 1 error in early phase clinical trials. \franz{When the drug is truly effective, success probabilities correspond to power; when the drug is not effective, success probabilities correspond to type 1 error.}}
\label{fig:results}
\end{sidewaysfigure}

\begin{sidewaysfigure}[ht]
\centering
\includegraphics[scale=0.47]{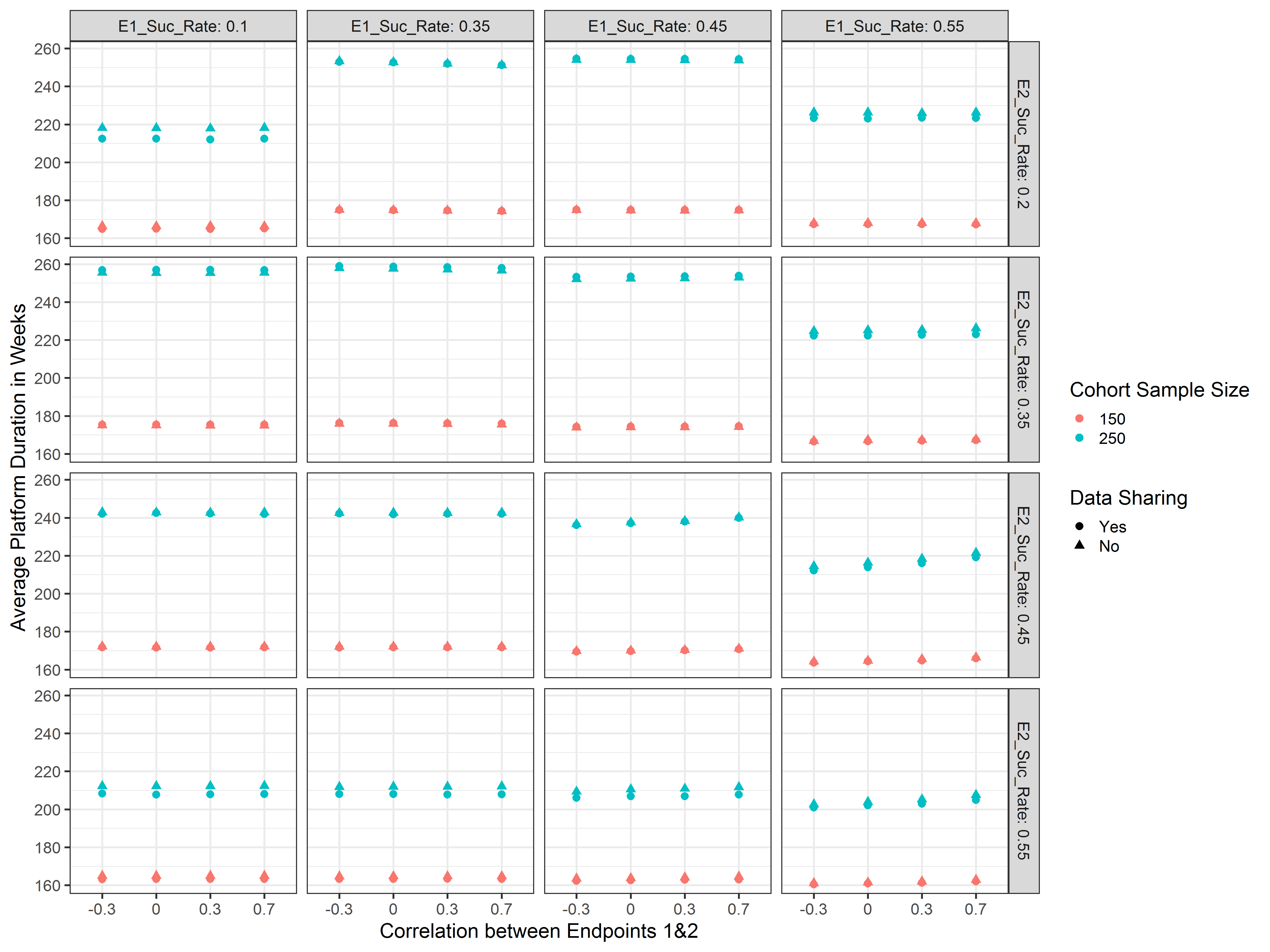}
\caption{Average platform trial duration in weeks with respect to the response rate for endpoint 1 (E1\_Suc\_Rate; columns), the response rate for endpoint 2 (E2\_Suc\_Rate; rows), the correlation between the two endpoints (x-axis), the type of data sharing used (point shape) and the planned cohort sample size per arm (colour).}
\label{fig:results2}
\end{sidewaysfigure}

\begin{sidewaysfigure}[ht]
\centering
\includegraphics[scale=0.38]{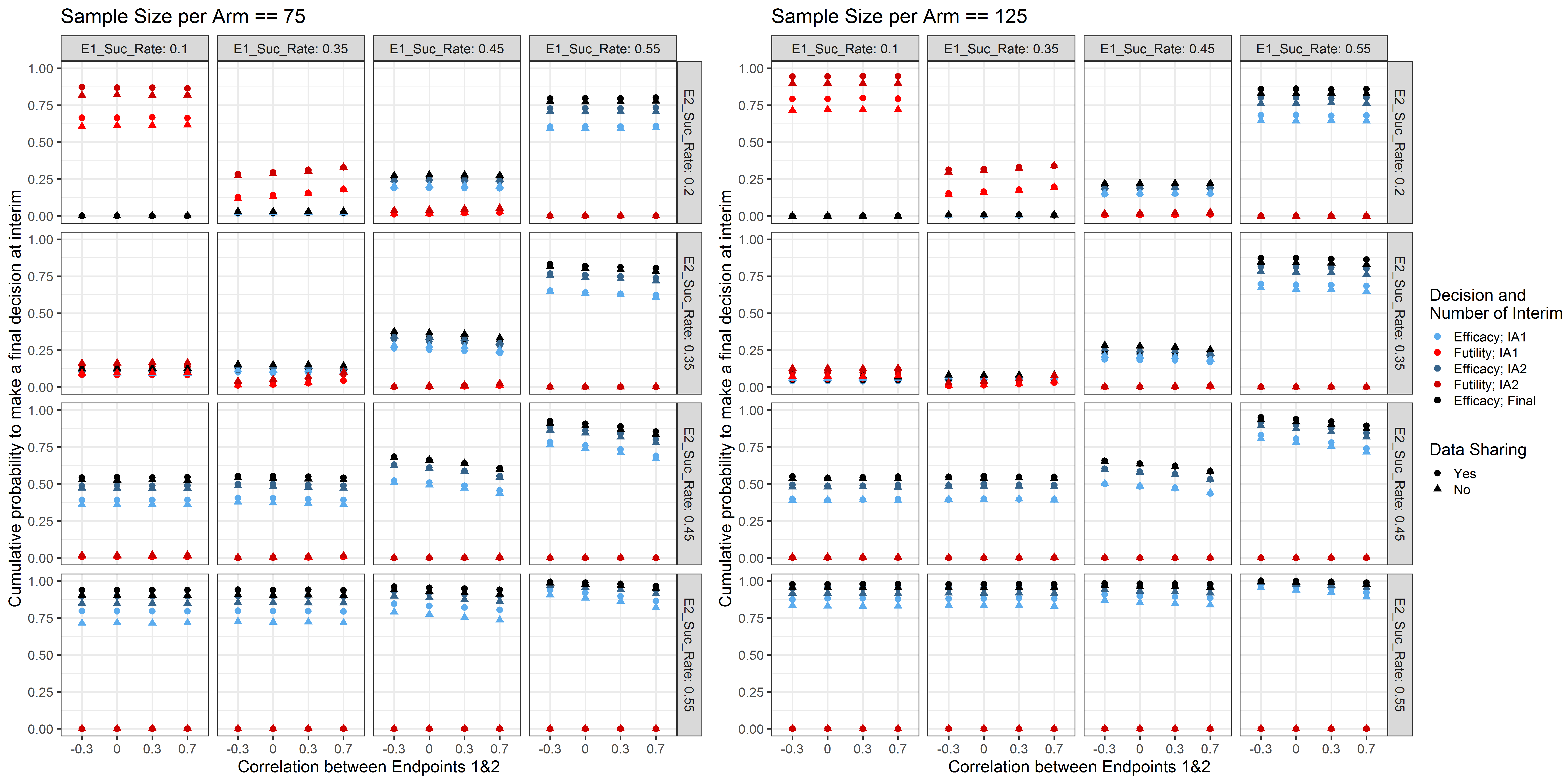}
\caption{Cumulative probabilities to make a decision early (i.e. making an early efficacy or futility decision either at the first or second interim analysis) with respect to the response rate for endpoint 1 (E1\_Suc\_Rate; columns), the response rate for endpoint 2 (E2\_Suc\_Rate; rows), the correlation between the two endpoints (x-axis), the type of data sharing used (point shape) and the planned sample size per cohort (left panel 150 and right panel 250).}
\label{fig:results3}
\end{sidewaysfigure}

Trial success probabilities with respect to the chosen simulation parameters are shown in Figure \ref{fig:results}. It becomes apparent that when the drug under investigation is not effective for either of the endpoints (i.e. responder rate of 10\% for endpoint 1 and 25\% for endpoint 2), the success probability (i.e. in this case type 1 error) is negligible (about 0.1\%) regardless of the sample size. On the other hand, when the drug is highly efficacious for either or both of the endpoints (i.e. responder rate of 55\% for both endpoints), the success probability (i.e. in this case power) is close to 1. When the treatment under investigation exhibits a responder rate of 35\% for one or both endpoints, the success probability is below 20\% and in case of a larger sample size (125 per arm) below 10\%. When the treatment under investigation is promisingly efficacious on both endpoints (i.e. responder rate of 45\%), the success probability is between 60-70\%, depending on the sample size and correlation. In terms of data sharing we observe the same pattern as with increased sample size - when the treatment is highly efficacious, the success probability increases, otherwise it decreases - this is a feature of the Bayesian decision rules (in frequentist analyses we would expect the type 1 error to be the same). \franz{As an example, when the sample size is 125 per arm, there is no correlation between the two endpoints and the success rate is 35\% for both endpoints, sharing data reduces the success probability (in this case corresponding to type 1 error) to 5\%, while the success probability is 8\% when not sharing data.} In general, we see a difference in success probabilities when only one of the two endpoints is reached - the success probability is higher if the treatment is efficacious on the Fibrosis endpoint. This is intended and consistent with our assumption that NASH resolution leads to Fibrosis improvement. Higher correlations between the two endpoints lead to reduced success probabilities - this is explained via the "OR" decision rule and the fact that outcomes are simulated using a latent bivariate normal distribution. This effect is most pronounced when the treatment is moderately efficacious for one or both of the endpoints (i.e. responder rate of 45\%). \newline
\newline
Average platform trial durations (i.e. time until a decision is made for the last investigational treatment) are shown in Figure \ref{fig:results2}. It becomes apparent that when the treatment is weakly efficacious for either or both of the endpoints (i.e. 35\% responder rate), trial duration is the longest (since it is unlikely that the treatment will be stopped for efficacy or futility at any of the interim analyses). For similar reasons, trial duration is decreased when the treatment is efficacious for neither of the endpoints and shortest when the treatment is highly efficacious for both of the endpoints. Generally, sharing concurrent data can lead to savings in trial duration of at most 2 weeks (sample size per arm 75; bottom left panel in Figure \ref{fig:results2}; 163 vs 165 weeks) or 6 weeks (sample size per arm 150; top left panel in Figure \ref{fig:results2}; 212 vs 218 weeks) compared to not sharing data (please note these numbers are also influenced by our assumption that new treatments would enter the platform every 24 weeks and the platform would necessarily run until five treatments are evaluated). Please note that since there is a lag of observing the final endpoints of 52 weeks, even if a decision is made early, this might not translate to savings in trial participants. In case the sample size per arm is 75, we observed no savings in terms of trial participants enrolled (i.e. always 750 trial participants are enrolled in the course of the trial). This is due to the assumed recruitment rate, which would lead to full recruitment in the time frame before the first interim analysis (a slower recruitment rate or more treatment arms investigated simultaneously might lead to savings). When the sample size per arm is 150, analogously to average trial duration, we observed savings in trial participants when treatments are overwhelmingly efficacious or futile (in the most extreme case of overwhelming efficacy on both endpoints and sharing data, approximately 382 out of 1500 trial participants were saved, compared to if no early stopping rules had been in place). \newline
\newline
Probabilities of stopping early with respect to treatment efficacy are shown in Figure \ref{fig:results3}. In case the treatment is not better than placebo, the futility rules eliminate approximately 60\% of treatments at the first interim analysis and approximately 80\% by the second interim analysis. This is true when the sample size per treatment arm is 75 and the probabilities increase further with increased sample size, i.e. 125. When the treatment is highly efficacious on both endpoints, the efficacy decision rules graduate most of the treatments at the first or second interim analysis. We also see that the futility interim decision rules eliminate more treatments that are only weakly effective on endpoint 1 and ineffective on endpoint 2 than if the reverse was true (this is a result of identical futility stopping rules while the standard-of-care response rate is lower for endpoint 1 than for endpoint 2).

\subsection{Impact of Bayesian Multi-Level Decision Rules}

\begin{sidewaysfigure}[ht]
\centering
\includegraphics[scale=0.38]{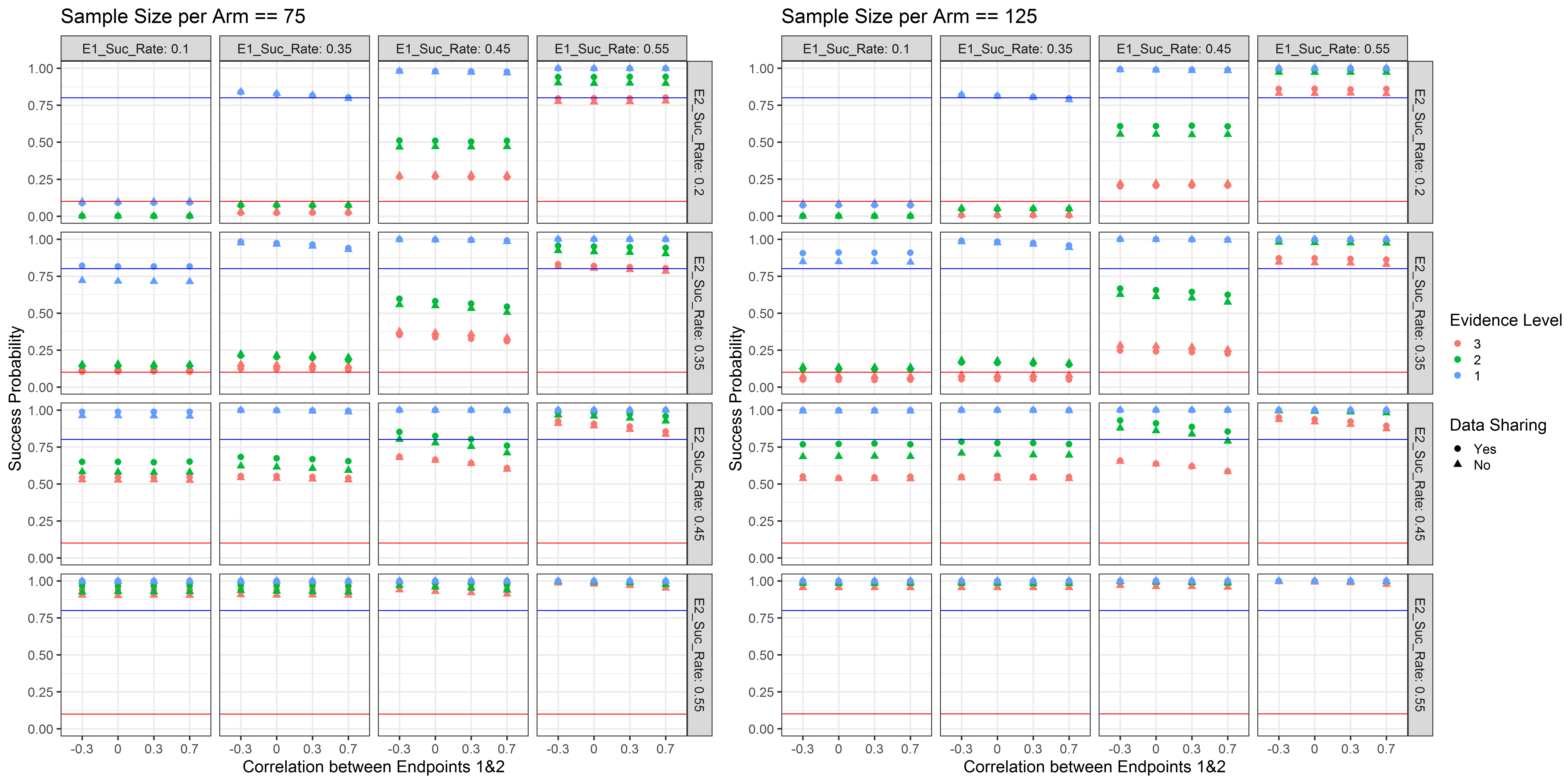}
\caption{Success probabilities for the treatment arm with respect to the response rate for endpoint 1 (E1\_Suc\_Rate; columns), the response rate for endpoint 2 (E2\_Suc\_Rate; rows), the correlation between the two endpoints (x-axis), the type of data sharing used (point shape), the level of evidence required (colour) and the planned sample size per cohort (left panel 150 and right panel 250). The blue horizontal line marks 80\% as a common target for the power and the red horizontal line marks 10\% as a common target for type 1 error in early phase clinical trials. Level of evidence required refers to how many of the Bayesian efficacy rules specified in section \ref{DR} need to simultaneously hold for a treatment to be declared efficacious.}
\label{fig:results4}
\end{sidewaysfigure}

Trial success probabilities with respect to the chosen simulation parameters as well as the level of evidence required in the Bayesian decision rules (see section \ref{DR}) are shown in Figure \ref{fig:results4}. It becomes apparent that when requiring only the lowest level of evidence (evidence level 1), type 1 error is controlled and for all investigated effect sizes there is a large success probability (which - depending on the target product profile - might be desired or not). When requiring a second level of evidence, success probabilities for effect sizes identified as insufficiently promising in our decision rule drop significantly, while success probabilities for large effect sizes stay large. When requiring all three levels of evidence, only treatments with a very large effect size are advanced with a high probability.

\section{Discussion}

For this exploratory phase 2b platform trial in NASH, a Bayesian framework has been chosen to incorporate the information from two endpoints (resolution of NASH without worsening of fibrosis (endpoint 1) and/or 1-stage fibrosis improvement without worsening of NASH (endpoint 2)) in the Bayesian decision rule. Based on the regulatory requirements, it has been decided that for the success criteria it is sufficient to demonstrate efficacy in either of the two endpoints. However, based on emerging phase 2b and phase 3 data for compounds under development, it is not sufficient to simply show superiority, instead there should be sufficient evidence that the effect sizes are large enough to show differentiation in order to graduate a treatment from phase 2b to phase 3. Therefore, the Bayesian decision rules have been extended to allow for different levels of evidence. Firstly, a high confidence is needed that the experimental treatment is better than control by any margin. Secondly, high confidence is required that the true effect is at least of moderate effect. Finally, some evidence is required that the true effect is relatively large and competitive with respect to the current landscape of compounds in the development pipeline. \newline
\newline
In frequentist trials, a study is powered to show superiority. The assessment whether the observed effect is relevant would be deferred to the lower bound of the confidence interval of the observed effect. Such a strategy could also be translated into a shifted hypothesis test using the duality between confidence intervals and frequentist tests. The proposed Bayesian framework allows a convenient way of combining the evidence from both endpoints and the extension to futility stopping rules in interim analyses. In this simulation study, we proposed that in order to declare success it is sufficient to demonstrate efficacy in either of the two endpoints, whereas both endpoints have to show insufficient efficacy to declare futility. In a frequentist trial, further multiplicity adjustments would have been needed for both repeated significance testing and testing two endpoints. We also investigated different ways in which two correlated binary endpoints can be simulated. The investigation of different correlations is critical for the trial's probability of success. Due to using an "OR" criterion (i.e. the treatment is graduated if either of the endpoints are met), a higher correlation will lead to decreased probabilities. Therefore, for planning purposes and determining sample sizes, it might be preferable to assume larger correlations, but ultimately this also depends on the chosen method to generate correlated binary endpoints. \newline
\newline
A draft of the proposed platform trial protocol including some simulation results were discussed with FDA in a critical path innovation (CPIM) meeting in January 2022 \citep{FDA_CPIM}. There were no objections to using Bayesian decision rules and it was acknowledged that in this phase 2b setting there is no need to correct for multiplicity on a platform level resulting from testing several compounds in the same trial. Regarding sharing of data, FDA supported the idea of using concurrent control data (defined as data of trial \franz{participants who were randomized to control arm in another cohort while randomization is ongoing in the cohort of interest who meet the inclusion/exclusion for all cohorts), but was opposed to using non-concurrent control data. Similar responses were received when discussing this NASH platform trial design with EMA in an Innovation Task Force (ITF) meeting in November 2022.} Within EU-PEARL, these results currently serve as a discussion foundation on how to choose the sample size and decision rules for the platform trial protocol as it seems that there is still some latitude with respect to type 1 error, especially considering that this is planned as a phase 2b design for decision making and not for registration purposes. \newline
\newline
\franz{One of the main advantages of platform trials is that they reduce the time and number of trial participants required to make a decision \citep{Meyer2020}. This is usually achieved both by operational and statistical efficiencies, such as multiple interim analyses and sharing data concurrently/non-concurrently across investigational cohorts. The impact of the interim analyses in reducing the duration of study and/or the number of participants will be lessened when the recruitment rate is fast relative to the time needed to observe the final endpoints (for example, if all of the participants needed for analysis in an investigational cohort are recruited in 3-6 months and the interim and final analyses are conducted at 12 months, there will be no savings in the number of participants entering the platform trial). This could lead to a large number of participants who have been randomized into a cohort in the trial, but have not yet had their primary endpoint observed when a decision is made to stop the cohort either due to superior or futile efficacy \cite{niewczas2019interim}. This problem becomes most evident in the trial design investigated here, when the sample size per treatmen arm is 75 with an accrual rate of 6 participants/week, even if a decision is made at one of the interim analyses to stop the cohort for superior efficacy, potentially there will be regulatory interest in seeing the treatment effect of the full cohort to provide further demonstration of the robustness of the efficacy observed at the interim analysis. If futility was demonstrated, then ethically you could prevent those randomized who have not reached week 52 from having an additional invasive liver biopsy. Therefore, initiatives to establish validated short-term endpoints in NASH based on biomarkers are critical. Savings in trial participants might be more pronounced when the recruitment rate is slower and/or more treatments are evaluated at the same time and or the sample size is larger. When the effect size is large, we observed time savings of around six weeks and up to 20-25\% reduction in the number of required participants when assuming 125 participants per treatment group and overwhelmingly superior or futile efficacy is observed along with the use of a concurrent control, indicating that even under the simple trial simulation assumptions studied the savings can be achieved under the ideal conditions. Further work is needed to determine how a Phase 2b NASH platform trial can achieve greater efficiency in the number of participants that need to be evaluated while making the best decisions in advancing those investigational treatments that have the potential of demonstrating transformative efficacy.}  \newline
\newline
Many extensions of this simulation study can be considered. First, more simulation parameters could be investigated - especially with respect to a range of accrual rates we would expect the savings in trial duration and participants to show meaningful change. The time between new treatments entering was set to 24 weeks and the maximum number of cohorts to 5, with two cohorts starting initially. If these assumptions were changed such that either more or less cohorts would be enrolling concurrently, differences in trial duration and success probabilities might be more pronounced for a concurrent control versus using cohort control groups only. Also, the use of non-concurrent controls could be investigated. We observed no significant changes in success probabilities when the sample size was increased beyond 75 trial participants per arm - therefore we believe no larger sample sizes are warranted based on the simulation parameters that have been evaluated to date. So far, it is assumed that trial participants are equally randomized between open cohorts and within cohorts between treatment arms. The use of response-adaptive-randomization might allow effective treatments to graduate faster - we did not investigate its use further, because we assumed identical treatment effects for all treatments within one platform trial. It was assumed that it is enough to show efficacy on one of the endpoints (i.e. "OR" decision rule). Future research could aim to show efficacy on both endpoints (i.e. "AND" decision rule). \newline
\franz{Both co-primary endpoints proposed for the this phase 2b platform trial have been discussed and agreed on by regulatory agencies, i.e. at an ITF meeting with EMA in November 2022 and a CPIM meeting with FDA in January 2022. While the resolution of steatosis or fibrosis improvement are clearly important endpoints, it may be scientifically interesting to evaluate disease progression to NASH cirrhosis. This might disclose that treatment regimens are unable to improve baseline condition, but possibly able to prevent disease progression. The only progression endpoint that is of regulatory interest is the one showing a progression from F2/F3 to F4. However, this is usually part of the clinical outcome composite endpoint evaluated in most NASH phase 3 trials and it would require longer follow-up than 12-18 months that is used in most Phase 2b NASH clinical trials. So when simulating phase 3 designs for NASH it should also incorporate the cumulative incidence of important clinical outcomes \citep{allen2022clinical, sanyal2021prospective} as the baseline risk and include clinically relevant outcomes  such as prevention of cirrhosis. However, the design of a phase III platform trial goes beyond the scope of this paper. The proposed platform trial is designed to incorporate phase 2b trials, but not phase 3 trials.} 
\newline
To conclude, we have found, based on our assumptions, that a NASH phase 2b platform study design employing Bayesian decision rules can demonstrate some time efficiencies when the effect sizes are large for either primary (histologic) endpoint, which is consistent with accelerating the development of transformational therapies. These time efficiencies would be in addition to those offered by a platform study in terms of accelerating start-up activities thereby creating a favorable proposition for drug developers, especially small biotechs. It is possible that once short-term (i.e., 12-24 week) biomarkers (alone or in combination) have sufficient data to predict effiacy for histological and/or clinical outcomes, the use of a platform design may become even more powerful using a phase 2a/2b seamless design, which could offer not just time savings but reduced sample size as well. However, for the moment based on current knowledge in the NASH field, the proposed design offers the benefits of potentially creating more opportunities for participants and overall reduced trial conduct time for developers leading to the main goal of EU-PEARL: providing tools for accelerating drug development with increased efficiency in a cross-sponsor approach that will ultimately benefit  patients.

\clearpage

\subsection*{Acknowledgements}

\franz{The authors are grateful to the NASH EU-PEARL investigators not included as authors for their work. The NASH EU-PEARL investigators are Nicholas DiProspero, Vlad Ratziu, Juan M. Pericàs, Mette Skalshøj Kjær, Quentin M. Anstee, Frank Tacke, Peter Mesenbrink, Jesús Rivera-Esteban, Franz Koenig, Elena Sena, Ramiro Manzano-Nunez, Joan Genescà, Raluca Pais, Leila Kara, Elias Laurin Meyer, Anna Duca, Timothy Kline, Anders Aaes-Jørgensen, Tania Balthaus, Natalie de Preville, Lingjiao Zhang, George Capuano, Salvatore Morello, Tobias Mielke, Sabina Hernandez Penna, and Martin Posch. Juan M. Pericàs led the NASH EU-PEARL investigators: juanmanuel.pericas@vallhebron.cat}


\footnotesize
\bibliography{ref}
\clearpage

\normalsize
\appendix

\section{Appendix}

\subsection{Detailed description of decision rules} \label{sec:app1}

\subsubsection{Verbal Description}

In the following, a verbal description of the decision rules presented in section \ref{DR} with respect to the analysis time point is given.

\paragraph{Interim Analysis 1}

Declare efficacy if either of the following two conditions is true:
    \begin{enumerate}
        \item 
            \begin{itemize}
                \item Posterior probability of at least 95\% that success rate of NASH resolution in treatment arm is by any margin larger than in SOC arm AND
                \item Posterior probability of at least 85\% that success rate of NASH resolution in treatment arm is by at least 30 percentage points larger than in SOC arm AND
                \item Posterior probability of at least 60\% that success rate of NASH resolution in treatment arm is by at least 40 percentage points larger than in SOC arm
            \end{itemize}
        \item 
            \begin{itemize}
                \item Posterior probability of at least 95\% that success rate of fibrosis improvement in treatment arm is by any margin larger than in SOC arm AND
                \item Posterior probability of at least 85\% that success rate of fibrosis improvement in treatment arm is by at least 17.5 percentage points larger than in SOC arm AND
                \item Posterior probability of at least 60\% that success rate of fibrosis improvement in treatment arm is by at least 25 percentage points larger than in SOC arm
            \end{itemize}
    \end{enumerate}
    
Declare futility if both of the following are true:
\begin{enumerate}
  \item Posterior probability of less than 20\% that success rate of NASH resolution in treatment arm is by at least 25 percentage points larger than in SOC arm
  \item Posterior probability of less than 20\% that success rate of fibrosis improvement in treatment arm is by at least 10 percentage points larger than in SOC arm
\end{enumerate}

If neither efficacy nor futility is declared, continue the trial.

\paragraph{Interim Analysis 2}

Declare efficacy if either of the following two conditions is true:
    \begin{enumerate}
        \item 
            \begin{itemize}
                \item Posterior probability of at least 95\% that success rate of NASH resolution in treatment arm is by any margin larger than in SOC arm AND
                \item Posterior probability of at least 85\% that success rate of NASH resolution in treatment arm is by at least 30 percentage points larger than in SOC arm AND
                \item Posterior probability of at least 60\% that success rate of NASH resolution in treatment arm is by at least 40 percentage points larger than in SOC arm
            \end{itemize}
        \item 
            \begin{itemize}
                \item Posterior probability of at least 95\% that success rate of fibrosis improvement in treatment arm is by any margin larger than in SOC arm AND
                \item Posterior probability of at least 85\% that success rate of fibrosis improvement in treatment arm is by at least 17.5 percentage points larger than in SOC arm AND
                \item Posterior probability of at least 60\% that success rate of fibrosis improvement in treatment arm is by at least 25 percentage points larger than in SOC arm
            \end{itemize}
    \end{enumerate}
    
Declare futility if both of the following are true:
\begin{enumerate}
  \item Posterior probability of less than 30\% that success rate of NASH resolution in treatment arm is by at least 25 percentage points larger than in SOC arm
  \item Posterior probability of less than 30\% that success rate of fibrosis improvement in treatment arm is by at least 10 percentage points larger than in SOC arm
\end{enumerate}

If neither efficacy nor futility is declared, continue the trial.

\paragraph{Final analysis}

Declare efficacy if either of the following two conditions is true:
    \begin{enumerate}
        \item 
            \begin{itemize}
                \item Posterior probability of at least 95\% that success rate of NASH resolution in treatment arm is larger than in SOC arm AND
                \item Posterior probability of at least 85\% that success rate of NASH resolution in treatment arm is by at least 30 percentage points larger than in SOC arm AND
                \item Posterior probability of at least 60\% that success rate of NASH resolution in treatment arm is by at least 40 percentage points larger than in SOC arm
            \end{itemize}
        \item 
            \begin{itemize}
                \item Posterior probability of at least 95\% that success rate of fibrosis improvement in treatment arm is larger than in SOC arm AND
                \item Posterior probability of at least 85\% that success rate of fibrosis improvement in treatment arm is by at least 17.5 percentage points larger than in SOC arm AND
                \item Posterior probability of at least 60\% that success rate of fibrosis improvement in treatment arm is by at least 25 percentage points larger than in SOC arm
            \end{itemize}
    \end{enumerate}

\subsubsection{Formal Definition} \label{sec:appDR}

Based on the decision rules given in equation \ref{eq:DR} in section \ref{DR}, the proposed multi-component decision rules for several endpoints and interim analyses can be generalized as follows:

\begin{align}
\begin{split}
\label{eq:DR2}
\text{GO, if } & (P(\pi_{E} > \pi_{S} + \delta_{k,1}^{G,T} | Data) > \gamma_{k,1}^{G,T}) \ \wedge \\
               & (P(\pi_{E} > \pi_{S} + \delta_{k,2}^{G,T} | Data) > \gamma_{k,2}^{G,T}) \ \wedge \\
               & \cdot \cdot \cdot \ \ \ \ \ \ \ \ \ \ \ \  \ \ \ \ \ \ \ \ \ \ \ \  \ \ \ \ \ \ \ \ \ \ \ \  \ \ \ \ \ \ \wedge \\
               & (P(\pi_{E} > \pi_{S} +  \delta_{k,l}^{G,T} | Data) > \gamma_{k,l}^{G,T}) \\
               \\
\text{STOP, if } & (P(\pi_{E} > \pi_{S} + \delta_{k,1}^{F,T} | Data) < \gamma_{k,1}^{F,T}) \ \vee \\
                 & (P(\pi_{E} > \pi_{S} + \delta_{k,2}^{F,T} | Data) < \gamma_{k,2}^{F,T}) \ \vee \\
                 & \cdot \cdot \cdot \ \ \ \ \ \ \ \ \ \ \  \ \ \ \ \ \ \ \ \ \ \ \  \ \ \ \ \ \ \ \ \ \ \ \  \ \ \ \ \ \ \vee \\
                 & (P(\pi_{E} > \pi_{S}  + \delta_{k,m}^{F,T} | Data) < \gamma_{k,m}^{F,T}) \\
                 \\
\end{split}
\end{align} 

whereby $\pi_S$ denotes the response rate in the standard-of-care arm, $\pi_E$ denotes the response rate in the experimental treatment arm, $T \in {1,2,...,N}$ denotes the analysis time point, subscript $k$ denotes the endpoint ($k \in \{1,...,K\}$) and subscripts $l$ and $m$ denote the possibility to have multiple decision rules at any given point in time. At interim ($T \in \{1,2,..N-1\}$), if neither a decision for early efficacy or futility is made, the cohort continues unchanged. At final ($T = N$), if the efficacy boundaries are not met, the cohort automatically stops for futility. The initial letters E or F in the superscript of the thresholds $\delta$ and $\gamma$ indicate if this boundary is used to stop for efficacy (G) or futility (F). Choosing, for example, $\gamma_{k,1}^{E,u} = 1 \ \forall u \in \{1, ... , k \}$ corresponds to not allowing early stopping for efficacy for endpoint $k$ at interim 1. If at any point in time both stopping for early efficacy and futility is allowed, parameters need to be chosen carefully such that GO and STOP and decisions are not simultaneously possible. Please note that the requirements in equation \ref{eq:DR2} refer to the evidence needed to declare efficacy or futility of a single endpoint, i.e. while we might have multiple efficacy requirements which need to be simultaneously fulfilled to declare a single endpoint efficacious, we advance the treatment if it is found to be efficacious in either of the two endpoints (analogously for futility decisions). \newline
\newline
In order to achieve the multi-level decision rules described in Table \ref{tab:hierarchy} and generalized in equation \ref{eq:DR2}, we set the following parameters for efficacy ($l=3$) and futility ($m=1$):

\begin{itemize}
\item $\delta_{k,1}^{G,T} = 0, \gamma_{k,1}^{G,T} = 0.95, \gamma_{k,2}^{G,T} = 0.85, \gamma_{k,3}^{G,T} = 0.60, \ \forall T \in \{1,2,3\} \ \forall k \in \{1,2\}$
\item $\delta_{1,2}^{G,T} = 0.30, \delta_{1,3}^{G,T} = 0.40, \delta_{2,2}^{G,T} = 0.175, \delta_{2,3}^{G,T} = 0.25, \ \forall T \in \{1,2,3\}$
\item $\gamma_{k,1}^{F,1} = 0.20, \gamma_{k,1}^{F,2} = 0.30, \gamma_{k,1}^{G,3} = 0, \ \forall k \in \{1,2\}$
\item $\delta_{1,1}^{F,T} = 0.25, \delta_{2,1}^{F,T} = 0.10 \ \forall T \in \{1,2,3\}$
\end{itemize}

\subsubsection{Visualization of Bayesian Decision Making based on Beta Posteriors} \label{sec:appBeta}

\begin{sidewaysfigure}[ht]
\centering
\includegraphics[scale=0.45]{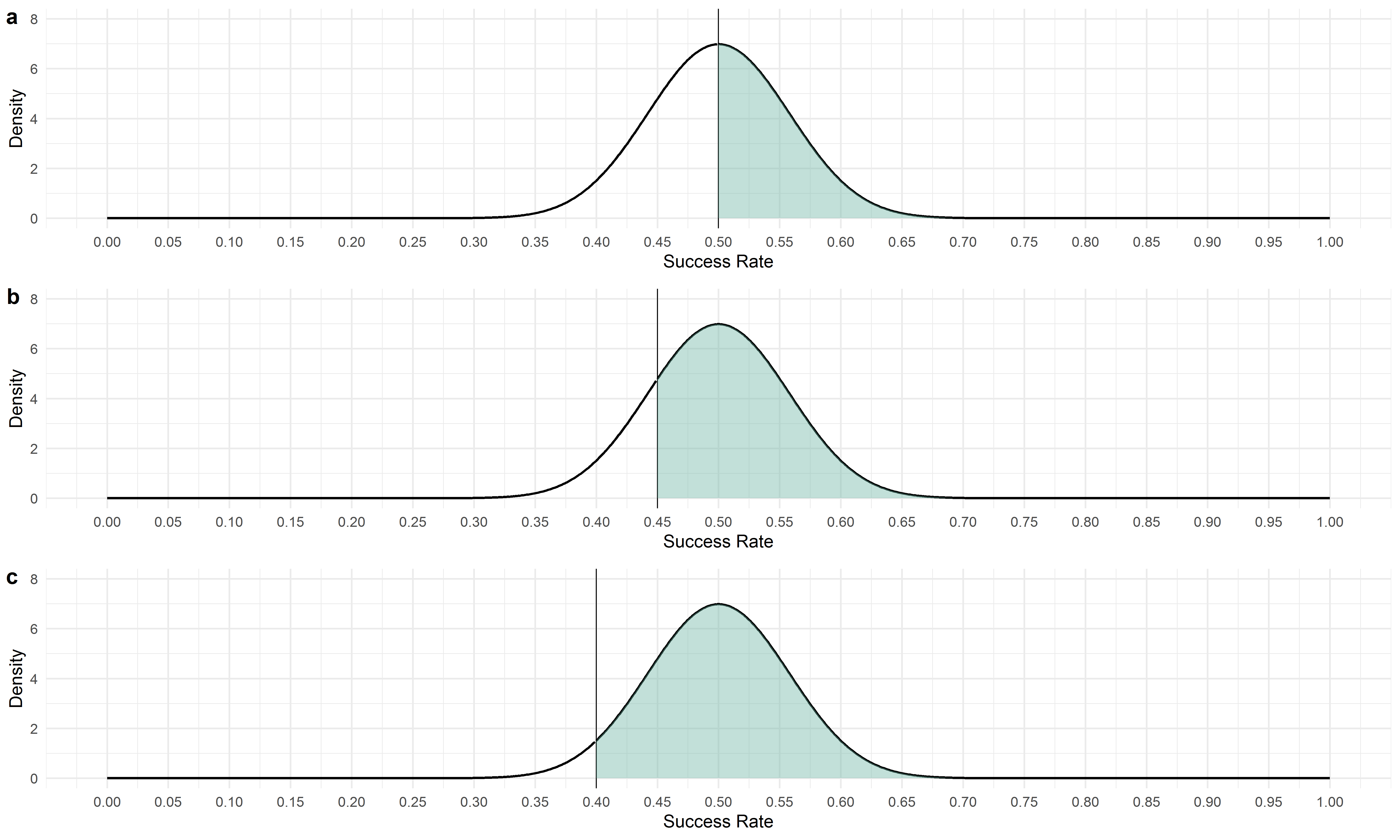}
\caption{Beta distributions corresponding to the posterior we would observe if a Beta(1,1) prior and a sample size of 75 was used and the observed response rate would equal 0.5. Panel a: The posterior probability for a success rate greater or equal to 0.5 is 50\%. If in our Bayesian decision rules we set a target of 0.5 and require a confidence of 50\%, for large sample sizes we will graduate compounds with true responder rate of 0.5 in 50\% of the cases, i.e. achieve a power of 50\%. Panel b: The posterior probability for a success rate greater or equal to 0.45 is 81\%. If in our Bayesian decision rules we set a target of 0.45 and require a confidence of 81\%, for large sample sizes we will graduate compounds with true responder rate of 0.5 in 50\% of the cases, i.e. achieve a power of 50\%. Panel c: The posterior probability for a success rate greater or equal to 0.40 is 96\%. If in our Bayesian decision rules we set a target of 0.40 and require a confidence of 96\%, for large sample sizes we will graduate compounds with true responder rate of 0.5 in 50\% of the cases, i.e. achieve a power of 50\%. In order to achieve larger power values, required confidences and target success rates need to be adapted.}
\label{fig:beta}
\end{sidewaysfigure}

The properties of Bayesian decision making using posterior probabilities as described in more detail in section \ref{DR} is highlighted in Figure \ref{fig:beta}.

\clearpage

    \subsection{Sampling Correlated Binary Endpoints} \label{CorrBin}

As mentioned previously, a successful clinical trial in NASH would investigate two correlated binary endpoints. This appendix is dedicated to establish the theory behind the sampling of correlated binary endpoints, which is used in our simulations in section \ref{sim_setup}. The proposed options are not meant to reflect the most efficient approaches (from a statistical or computational perspective), but rather the most realistic in terms of an interdisciplinary collaboration between clinicians and statisticians, whereby the role of the latter is to translate the received information as accurately as possible into simulation programs.  \newline
\newline
Let us assume during the course of a clinical trial investigating one or more treatments, we observe data on two correlated binary outcomes for every trial participant, $S$ and $L$. This could be the case if e.g. $L$ is a long-term endpoint and $S$ is a short-term endpoint which might be used as a surrogate for $L$. Another example would be if $S$ and $L$ are co-primary endpoints, as is the case in the trial design under investigation where S is NASH resolution (i.e. endpoint 1) and L is fibrosis improvement (i.e. endpoint 2). For the remainder of this section however, for reasons of better readability, we will refer to the two endpoints as "short-term" (S) and "long-term" (L) endpoint. For any given trial participant $i$, the four possible events which we could observe and their respective probabilities are $P(S = 0, L = 0) = p^i_{00}, P(S = 1, L = 0) = p^i_{10}, P(S = 0, L = 1) = p^i_{01}$ and $P(S = 1, L = 1) = p^i_{11}$. We assume that $p_{sl}^i$ is fully characterized by trial participant $i$’s covariates $x_i$. If treatment is the only covariate, $p_{sl}^i=p_{sl}^{k_i}$ where $k_i=k$ denotes treatment. Therefore, for the remainder of this section we drop the superscript $i$ in favor of a superscript $k$ on these probabilities, such that e.g. $p^k_{00}$ denotes the probability to have no success on either endpoint for all trial participants receiving treatment $k$. Hence, $(S_i, L_i)$ follows a bivariate Bernoulli distribution with expected value $\mathbf{p_i}^k$, covariance 
$p^k_{11} * p^k_{00} - p^k_{01} * p^k_{10}$ and correlation (Pearson $\phi$): $$
\phi^k = \frac{p^k_{11} * p^k_{00} - p^k_{01} * p^k_{10}}{\sqrt{p^k_{0\cdot} * p^k_{1\cdot} * p^k_{\cdot0} * p^k_{\cdot1}}} = \frac{p^k_{11} - p^k_{1\cdot} * p^k_{\cdot1}}{\sqrt{p^k_{0\cdot} * p^k_{1\cdot} * p^k_{\cdot0} * p^k_{\cdot1}}}
$$
where $\mathbf{p_i}^k$ denotes vector of marginal success probabilities for $S$ and $L$ for trial participant $i$ (i.e. we only consider two different such vectors $\mathbf{p_i}^k$ if treatment is the only covariate) \citep{marshall1985family}. For an overview see Table \ref{tab:2x2}. For a recent article discussing incorporation of both short-term and long-term binary data at interim, see \citet{niewczas2019interim}, where simulations were based on the \textbf{R} packages \textbf{mvtnorm} and \textbf{psych}. Generalizations of this problem for more than two dimensions are discussed in \citet{jiang2021set}.

\begin{table}[ht]
\centering
\caption{2x2 crosstable of possible short-term (S) and long-term (L) outcomes and their respective probabilities, as well as the marginal probabilities, with respect to treatment (denoted by superscript $k$).}
\label{tab:2x2}
\begin{tabular}{ c| c c | c}
 S/L & 0 & 1 & \\ 
 \hline
 0 & $p^k_{00}$ & $p^k_{01}$ & $p^k_{0\cdot}$ \\  
 1 & $p^k_{10}$ & $p^k_{11}$ & $p^k_{1\cdot}$ \\
 \hline
   & $p^k_{\cdot0}$ & $p^k_{\cdot1}$ & 1\\
\end{tabular}
\end{table}

Different diagnostic and predictive properties that arise are:

\begin{itemize}
    \item Sensitivity of the short-term endpoint in predicting the long-term endpoint ($sens^k_{SL}$): $\frac{p^k_{11}}{p^k_{1\cdot}}$
    \item Specificity of the short-term endpoint in predicting the long-term endpoint ($spec^k_{SL}$): $\frac{p^k_{00}}{p^k_{0\cdot}}$
    \item Sensitivity of the long-term endpoint in predicting the short-term endpoint ($sens^k_{LS}$): $\frac{p^k_{11}}{p^k_{\cdot1}}$
    \item Specificity of the long-term endpoint in predicting the short-term endpoint ($spec^k_{LS}$): $\frac{p^k_{00}}{p^k_{\cdot0}}$
\end{itemize}
\noindent
There are multiple ways to specify the required probabilities in Table \ref{tab:2x2}. We will now explore a few options in more detail which as mentioned previously focus on practical applicability in a setting of interdisciplinary collaboration and not necessarily on statistical or mathematical efficiency. Except for the method in section \ref{sec:MNT}, all of the explored methods are reparametrizations of $\mathbf{p}= (p_{00}, p_{10}, p_{01}, p_{11})$ with the constraints $p_{ij} \in [0,1]$ and $\sum_{ij} p_{ij}=1 $ (e.g. bijective functions which map $(p_{00}, p_{01}, p_{10})$ onto another set of three parameters). 

\subsubsection{Direct specification}

The easiest and most straightforward way to specify the joint distribution is by fixing the four probabilities $p^k_{00}, p^k_{10}, p^k_{10}$ and $p^k_{11} = 1 - (p^k_{00} + p^k_{10} + p^k_{01})$. We can then calculate the correlation and diagnostic and predictive properties as described in the previous section. A drawback of this method is that likely neither any diagnostic/predictive properties nor the correlation are the same for both treatments unless the same probabilities are specified, which makes this approach rather unintuitive and results difficult to communicate.

\subsubsection{Implicit specification via Sensitivity \& Specificity or PPV \& NPV}

Next, let us re-consider Table \ref{tab:2x2} and note that there are various different ways of picking at least 3 of the unknown variables to fully characterize the bivariate Bernoulli distribution. In the previous section, we chose three unknown parameters from the "inside" of the Table (the fourth directly followed), but in this subsection we explore the option of specifying one marginal probability (i.e. either $p^k_{1\cdot}$ or $p^k_{\cdot1}$) and the diagnostic and predictive properties of this outcome in predicting the other outcome (i.e. in case of specifying $p^k_{1\cdot}$, we additionally specify $sens^k_{SL}$ and $spec^k_{SL}$). This is possible without any constraints on the chosen probabilities.  This approach might make sense if, for example, $S$ is a surrogate endpoint for $L$.  Note that specifying one marginal probability and the diagnostic and predictive properties of the other outcome in predicting this outcome (i.e. in case of specifying $p^k_{1\cdot}$, additionally specifying $sens^k_{LS}$ and $spec^k_{LS}$) is possible only under heavy constraints on the chosen triplet of values and most combinations of values are not attainable (for more information see Figure \ref{fig:corsensspez}). Another drawback is that in most clinical examples, it might not make sense to require $p^k_{1\cdot}$ and $sens^k_{SL}$ and $spec^k_{SL}$ as input parameters for a simulation study, but rather $p^k_{\cdot1}$ and $sens^k_{SL}$ and $spec^k_{SL}$, which, as discussed before, leads to invalid probabilities for most combinations of these three parameters. \newline
\newline
The required transformations are reported exemplary for specifying $p^k_{1\cdot}$, $sens^k_{SL}$ and $spec^k_{SL}$. From $sens^k_{SL} = \frac{p^k_{11}}{p^k_{1\cdot}}$ and $spec^k_{SL} = \frac{p^k_{00}}{p^k_{0\cdot}}$ it follows that $p^k_{11} = sens^k_{SL} * p^k_{1\cdot}$ and $p^k_{00} = spec^k_{SL} * (1 - p^k_{1\cdot})$. Subsequently, $p^k_{10} = p^k_{1\cdot} - p^k_{11}$ and $p^k_{01} = (1 - p^k_{1\cdot}) - p^k_{00}$. Another drawback of this method is that the correlation and diagnostic and predictive properties of $S$ in predicting $L$ will likely differ between treatments. 

\subsubsection{Implicit specification via Correlation}

In this subsection we explore the option of specifying $p^k_{1\cdot}$, $p^k_{\cdot1}$ and $\phi^k$. From the second formula for $\phi^k$ provided in the previous section, it follows that  

$$
p^k_{11} = \phi^k*\sqrt{p^k_{0\cdot} * p^k_{1\cdot} * p^k_{\cdot0} * p^k_{\cdot1}} + p^k_{1\cdot} * p^k_{\cdot1}
$$
\noindent
As other authors have noted \citep{sozu2010sample}, after fixing $p^k_{\cdot1}$ and $p^k_{1\cdot}$, $\phi^k$ cannot be chosen freely anymore and is bounded below by 

$$
max\Bigg( -\sqrt{\frac{p^k_{1\cdot} * p^k_{\cdot1}}{(1-p^k_{1\cdot})(1-p^k_{\cdot1})}}, -\sqrt{\frac{(1-p^k_{1\cdot})(1-p^k_{\cdot1})}{p^k_{1\cdot} * p^k_{\cdot1}}} \Bigg) 
$$

and above by 

$$
min\Bigg( \sqrt{\frac{p^k_{1\cdot} * (1-p^k_{\cdot1})}{(1-p^k_{1\cdot})*p^k_{\cdot1}}}, \sqrt{\frac{(1-p^k_{1\cdot})*p^k_{\cdot1}}{p^k_{1\cdot} * (1-p^k_{\cdot1})}} \Bigg) 
$$
\noindent
Finally, $p^k_{01} = p^k_{\cdot1} - p^k_{11}$, $p^k_{10} = p^k_{1\cdot} - p^k_{11}$ and $p^k_{00} = 1 - (p^k_{11} + p^k_{01} + p^k_{10})$ can simply be derived. \newline
\newline
The constraints on the triplet $ (p^k_{1\cdot}, p^k_{\cdot1}, \phi^k) $ have two major impacts in the scenario of simulating correlated binary outcomes for different treatments: 1) If we assume the correlation to be the same across all treatments, it might happen that for some treatments we derive valid probabilities and for some treatments we do not derive valid probabilities. 2) For any chosen pair $ (p^k_{1\cdot}, p^k_{\cdot1})$, we are unable to investigate the whole range of correlations. Frequently statisticians will receive the following information, based on which the simulations should be performed: "We know $(p^k_{1\cdot}$ and $p^k_{\cdot1})$ and the correlation". While this method would technically be the correct approach to take, it is very hard to keep track of the restrictions on the parameters (since one of the parameters has a support which depends on the values of the other two) and they are also difficult to communicate.

\subsubsection{Specification via bivariate normal distribution} \label{sec:MNT}

\begin{figure}[!ht]
\centering
\includegraphics[scale=0.35]{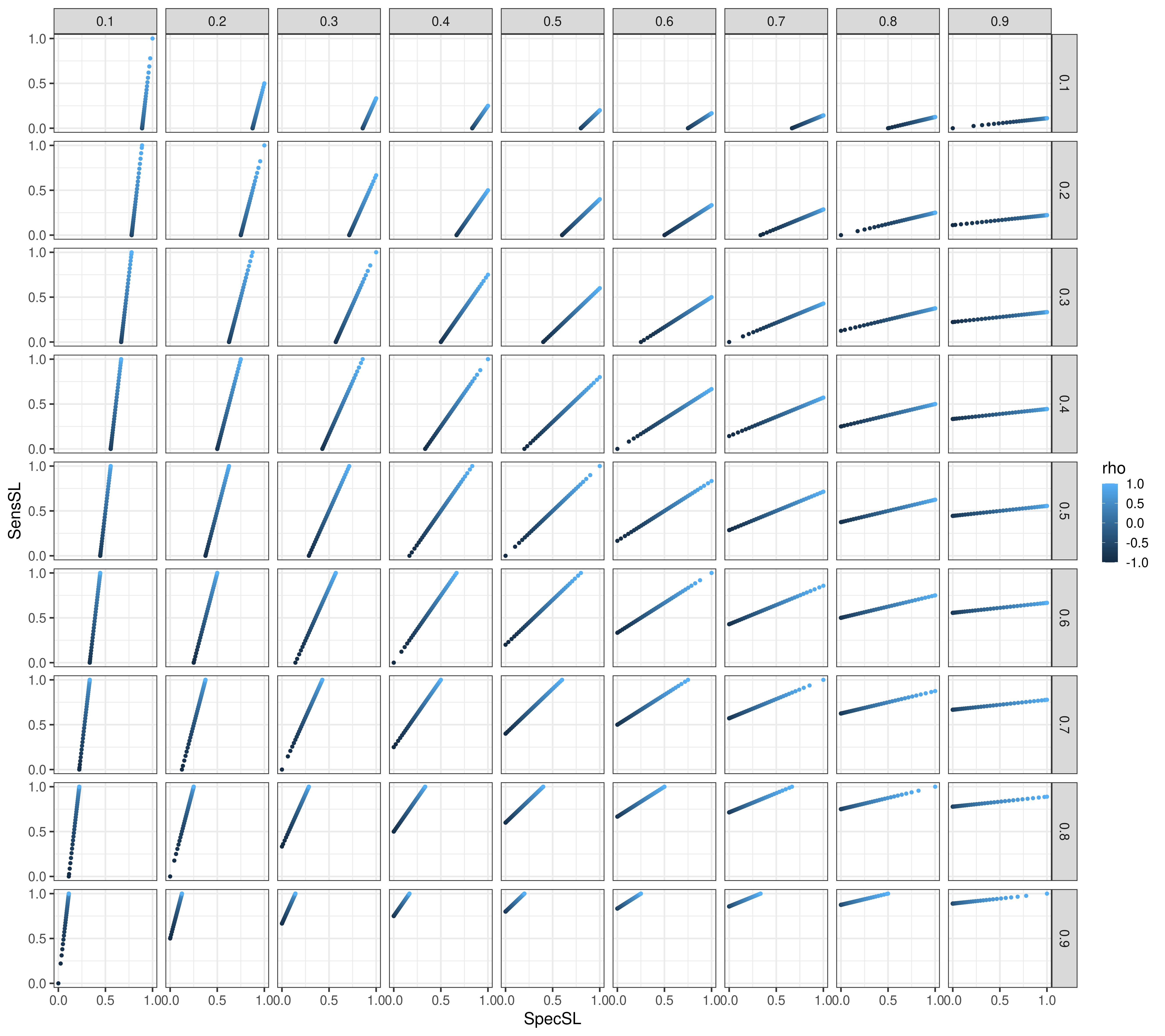}
\caption{$sens_{SL}$ and $spec_{SL}$ with respect to $p^k_{1\cdot}$ (columns), $p^k_{\cdot1}$ (rows) and $\rho$ (shade of color). For scenarios $p^k_{1\cdot} < p^k_{\cdot1}$, there is an upper $u_{spec} < 1$ bound for $spec_{SL}$. For scenarios $p^k_{1\cdot} > p^k_{\cdot1}$, there is an upper $u_{sens} < 1$ bound for $sens_{SL}$. For scenarios $p^k_{1\cdot} = p^k_{\cdot1}$ either $sens_{SL}$ ($p^k_{1\cdot} = p^k_{\cdot1} \leq 0.5$) or $spec_{SL}$ ($p^k_{1\cdot} = p^k_{\cdot1} \geq 0.5$) can take any values between 0 and 1. If the matrix of figures was transposed, we would see $sens_{LS}$ and $spec_{LS}$ instead of $sens_{SL}$ and $spec_{SL}$. Please note that in the Figure the label ``rho" is used for $\rho$.}
\label{fig:corsensspez}
\end{figure}

Finally, the required probabilities can be derived using a latent variable approach from a bivariate normal distribution. In this approach, we would usually like to fix $p^k_{\cdot1}$ (and thereby $p^k_{\cdot0}$) and $p^k_{1\cdot}$ (and thereby $p^k_{0\cdot}$), i.e. the short-term and long-term response rates of treatment $k$. Furthermore, we would like to specify a naive correlation $\rho$ for the two outcomes, i.e. we need to specify a triplet $(p^k_{\cdot1}, p^k_{1\cdot}, \rho)$. We then define the joint probabilities as follows:
$$
p^k_{00} = \int_{-\infty}^{p^k_{0\cdot}} \int_{-\infty}^{p^k_{\cdot0}} f(\vec{x}, \vec{0}, \left[ {\begin{array}{cc}
    1 & \rho \\
    \rho & 1 \\
  \end{array} } \right]) dx_1dx_2
$$

$$
p^k_{10} = \int_{p^k_{0\cdot}}^{\infty} \int_{-\infty}^{p^k_{\cdot0}} f(\vec{x}, \vec{0}, \left[ {\begin{array}{cc}
    1 & \rho \\
    \rho & 1 \\
  \end{array} } \right]) dx_1dx_2
$$

$$
p^k_{01} = \int_{-\infty}^{p^k_{0\cdot}} \int^{\infty}_{p^k_{\cdot0}} f(\vec{x}, \vec{0}, \left[ {\begin{array}{cc}
    1 & \rho \\
    \rho & 1 \\
  \end{array} } \right]) dx_1dx_2
$$

$$
p^k_{11} = \int_{p^k_{0\cdot}}^{\infty} \int^{\infty}_{p^k_{\cdot0}} f(\vec{x}, \vec{0}, \left[ {\begin{array}{cc}
    1 & \rho \\
    \rho & 1 \\
  \end{array} } \right]) dx_1dx_2
$$
\noindent
whereby $f(\vec{x}, \vec{\mu}, \Sigma)$ is the probability density function of the bivariate normal distribution with mean $\mu$, covariance matrix $\Sigma$ and $\vec{x}$ = $(x_1, x_2)$. It should be obvious that these four probabilities sum up to 1. This basically corresponds to splitting the bivariate normal distribution into 4 new quadrants and setting the probabilities equal to the probability mass in each of these four quadrants.  \newline
\newline
The most obvious drawback of this methods is that not all combinations of $\mathbf{p_i}$ are attainable this way (e.g. $(0.33, 0.34, 0.33, 0)$ is impossible). Another drawback of allowing the specification of triplets $(p^k_{\cdot1}, p^k_{1\cdot}, \rho)$ is that the diagnostic and predictive properties differ for different treatments (if they have different response rates and equal correlations). In fact, for a given pair of response rates $(p^1_{1\cdot}, p^1_{\cdot1})$, the sensitivity and specificity are a function of the correlation $\rho$. Therefore, there are constraints regarding achievable sensitivity and specificity for a given set of response rates $(p^k_{\cdot1}, p^k_{1\cdot})$. Figure \ref{fig:corsensspez} shows the achieved sensitivity and specificity for different ranges of $p^k_{\cdot1}$ and $p^k_{1\cdot}$. Another drawback of this method is the non-linear relationship between the specified $\rho$ and the actual correlation of the binary endpoints, $\phi$, which also depends on $p^k_{\cdot1}$ and $p^k_{1\cdot}$. In practise, this means that in most cases (as in our simulation study) we will think of $\rho$ as the correlation, when in fact the actual correlation $\phi$ between the binary endpoints might differ substantially. See Figure \ref{fig:phirho} in the appendix for more details. \newline
\newline
As demonstrated in the previous sections, no single specification option comes without limitations. For this simulation study, we chose to sample trial participants' correlated binary endpoints via the bivariate normal distribution, because requiring marginal success probabilities and a "naive" correlation seemed like the best trade-off between being unproblematic in terms of constraints, being easy to implement in the software and finally - and most importantly - easy to communicate to clinical teams.

\begin{figure}[!ht]
\centering
\includegraphics[scale=0.35]{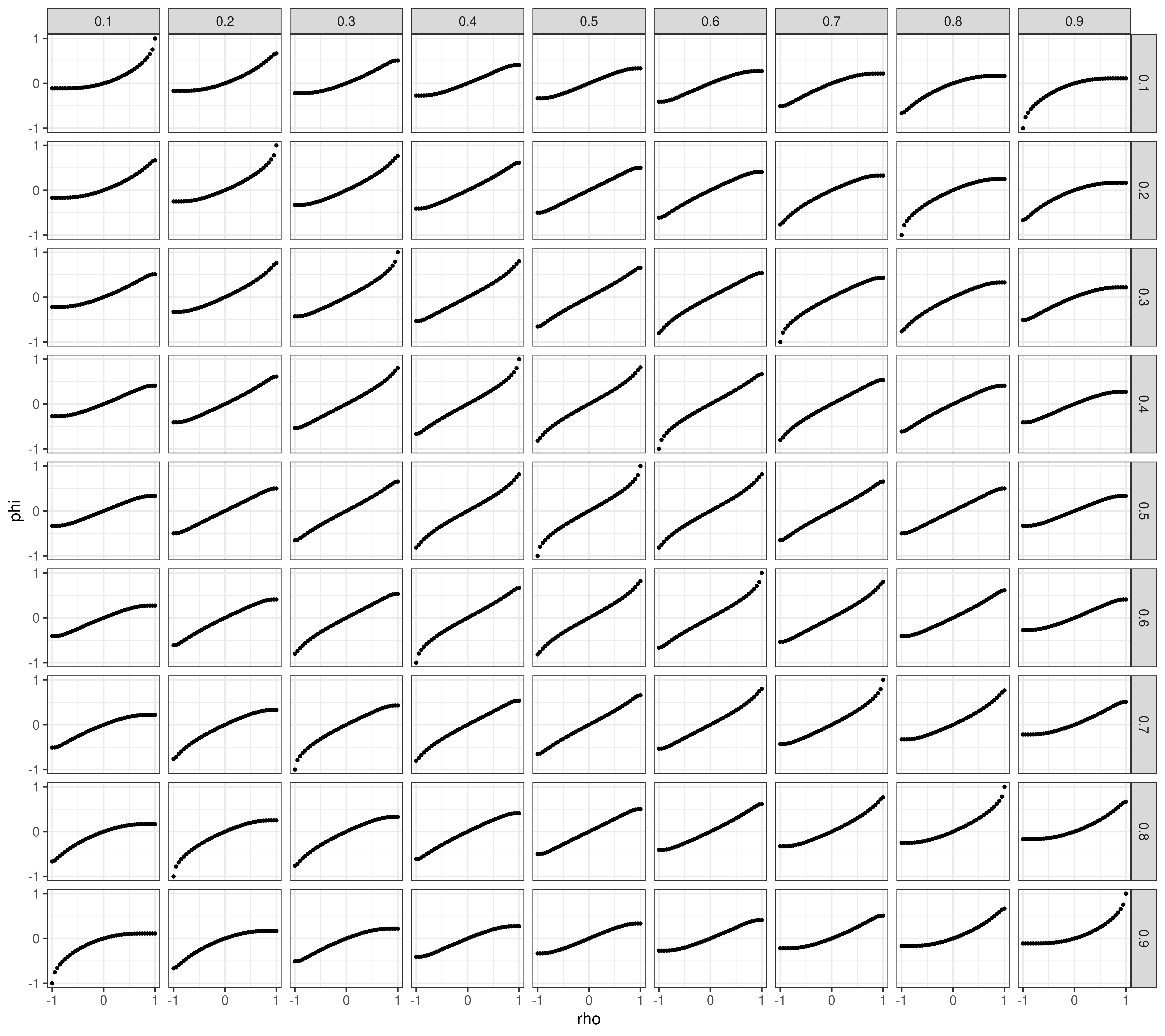}
\caption{Relationship between specified correlation of the two continuous endpoints $\rho^k$ prior to dichotomization and achieved correlation $\phi^k$ of the two binary endpoints after dichotomization. Only when $p^k_{1\cdot} = p^k_{\cdot1} = 0.5$ can $\phi^k \in [0,1]$ be achieved, otherwise it is bounded above and/or below (see paragraph "Correlation" in section "Implicit specification" for more details on the bounds). The more $p^k_{1\cdot}$ and $p^k_{\cdot1}$ differ, the closer either the upper or lower bound of $\phi^k$ is to 0. Please note that in the Figure the labels ``phi" and ``rho" are used for $\phi$ and $\rho$.}
\label{fig:phirho}
\end{figure}

\end{document}